\documentclass{nature}

\newcounter{firstbib}

\usepackage{graphicx}
\usepackage{epsfig}
\usepackage[font={small}]{caption}
\usepackage{tablefootnote}

\def\gtorder{\mathrel{\raise.3ex\hbox{$>$}\mkern-14mu
             \lower0.6ex\hbox{$\sim$}}}
\def\ltorder{\mathrel{\raise.3ex\hbox{$<$}\mkern-14mu
             \lower0.6ex\hbox{$\sim$}}}

\usepackage{amssymb}
\usepackage{amsmath}
\usepackage{tablefootnote}

\newcommand{\aap}{Astron. Astrophys.}
\newcommand{\araa}{Ann. Rev. Astron. Astrophys.}
\newcommand{\apj}{Astrophys. J.}

\newcommand{\apjl}{Astrophys. J. Letters}
\newcommand{\apjs}{Astrophys. J. Supplements}
\newcommand{\nat}{Nature}
\newcommand{\pasp}{Publications of the Astronomical Society of the Pacific}

\newcommand{\mnras}{Mon. Not. R. Astron. Soc.}

\newcommand{\prd}{Phys.R. D}
\newcommand{\apss}{Ap\&SS}  

\title{The Superluminous Transient ASASSN-15lh as a Tidal Disruption Event from a Kerr Black Hole}

\author{G. Leloudas$^{1,2}$, M. Fraser$^3$, N.~C. Stone$^4$, S. van Velzen$^5$, P.~G. Jonker$^{6,7}$, I. Arcavi$^{8,9}$, C.~Fremling$^{10}$,
J.~R.~Maund$^{11}$, S.~J. Smartt$^{12}$, T. Kr\"uhler$^{13}$, J.~C.~A. Miller-Jones$^{14}$, P.~M.~Vreeswijk$^1$, A. Gal-Yam$^1$, 
P.~A. Mazzali$^{15,16}$, A. De Cia$^{17}$, D.~A. Howell$^{8,18}$, C. Inserra$^{12}$, F. Patat$^{17}$, A. de Ugarte Postigo$^{19,2}$,
O. Yaron$^1$, C. Ashall$^{15}$, I. Bar$^1$, H. Campbell$^{3,20}$, T.-W. Chen$^{13}$, M. Childress$^{21}$, N.~Elias-Rosa$^{22}$, J. Harmanen$^{23}$, 
G. Hosseinzadeh$^{8,18}$, J. Johansson$^1$, T. Kangas$^{23}$, E. Kankare$^{12}$, S. Kim$^{24}$, H. Kuncarayakti$^{25,26}$, 
J. Lyman$^{27}$, M.~R. Magee$^{12}$, K. Maguire$^{12}$, D. Malesani$^2$, S. Mattila$^{23,28,3}$, C.~V. McCully$^{8,18}$, M. Nicholl$^{29}$, 
S. Prentice{$^{15}$}, C. Romero-Ca\~nizales$^{24,25}$, S. Schulze$^{24,25}$, K.~W.~Smith$^{12}$, J. Sollerman$^{10}$, M. Sullivan$^{21}$, 
B.~E. Tucker$^{30,31}$, S. Valenti$^{32}$, J.~C. Wheeler$^{33}$, D.~R. Young$^{12}$}
\begin{document}

\maketitle

\begin{affiliations}
   \item Department of Particle Physics and Astrophysics, Weizmann Institute of Science, Rehovot 7610001, Israel 
   \item Dark Cosmology Centre, Niels Bohr Institute, University of Copenhagen, Juliane Maries vej 30, 2100 Copenhagen, Denmark
   \item Institute of Astronomy, University of Cambridge, Madingley Road, Cambridge CB3 0HA, UK
   \item Columbia Astrophysics Laboratory, Columbia University, New York, NY, 10027, USA
   \item Department of Physics and Astronomy, The Johns Hopkins University, Baltimore, MD, 21218, USA
   \item SRON, Netherlands Institute for Space Research, Sorbonnelaan 2, 3584 CA, Utrecht, The Netherlands
   \item Department of Astrophysics/IMAPP, Radboud University Nijmegen, P.O. Box 9010, 6500 GL Nijmegen, The Netherlands
   \item Las Cumbres Observatory Global Telescope Network, 6740 Cortona Dr., Suite 102 Goleta, CA 93117, USA
   \item Kavli Institute for Theoretical Physics, University of California, Santa Barbara, CA 93106, USA
   \item Department of Astronomy, The Oskar Klein Center, Stockholm University, AlbaNova, 10691, Stockholm, Sweden
   \item Department of Physics and Astronomy, The University of Sheffield, Hicks Building, Hounsfield Road, Sheffield, S3 7RH, UK
   \item Astrophysics Research Centre, School of Mathematics and Physics, Queen's University Belfast, Belfast BT7 1NN, UK
   \item Max-Planck-Institut f\"{u}r extraterrestrische Physik, Giessenbachstra\ss e, 85748 Garching, Germany
   \item International Centre for Radio Astronomy Research, Curtin University, GPO Box U1987, Perth, WA 6845, Australia
   \item Astrophysics Research Institute, Liverpool John Moores University, IC2, Liverpool Science Park, 146 Brownlow Hill, Liverpool L3 5RF, UK
   \item Max-Planck Institut f\"ur Astrophysik, Karl-Schwarzschild-Str. 1, 85748 Garching b. M\"unchen, Germany
   \item European Southern Observatory, Karl-Schwarzschild-Strasse 2, 85748 Garching, Germany
   \item Department of Physics, University of California Santa Barbara, Santa Barbara, CA 93117, USA
   \item Instituto de Astrof\' isica de Andaluc\' ia (IAA-CSIC), Glorieta de la Astronom\' ia s/n, E-18008, Granada, Spain
   \item Department of Physics, University of Surrey, Guildford, GU2 7XH, Surrey, UK
   \item Department of Physics \& Astronomy, University of Southampton, Southampton, Hampshire, SO17 1BJ, UK
   \item INAF - Osservatorio Astronomico di Padova, Vicolo dellÕOsservatorio 5, 35122 Padova, Italy
   \item Tuorla Observatory, Department of Physics and Astronomy, University of Turku, V\"ais\"al\"antie 20, FI-21500 Piikki\"o, Finland
   \item Instituto de Astrof\'isica, Facultad de F\'isica, Pontificia Universidad Cat\'olica de Chile, Vicu\~{n}a Mackenna 4860, 7820436 Macul, Santiago, Chile
   \item Millennium Institute of Astrophysics, Santiago, Chile
   \item Departamento de Astronom\'ia, Universidad de Chile, Casilla 36-D, Santiago, Chile
   \item Department of Physics, University of Warwick, Coventry CV4 7AL, UK
   \item Finnish Centre for Astronomy with ESO (FINCA), University of Turku, V\"ais\"al\"antie 20, FI-21500 Piikki\"o, Finland
   \item Harvard-Smithsonian Center for Astrophysics, 60 Garden Street, Cambridge, Massachusetts 02138, USA
   \item The Research School of Astronomy and Astrophysics, Mount Stromlo Observatory, Australian National University, Canberra, ACT 2611, Australia.
   \item ARC Centre of Excellence for All-sky Astrophysics (CAASTRO), Australia
   \item Department of Physics, University of California, Davis, CA 95616, USA
   \item Department of Astronomy, University of Texas at Austin, Austin, TX 78712, USA
\end{affiliations}


\clearpage

\begin{abstract}
When a star passes within the tidal radius of a supermassive black hole, it will be torn apart\cite{Rees88}. 
For a star with the mass of the Sun ($M_\odot$) and a non-spinning black hole with a mass $<10^8 M_\odot$, the tidal radius lies outside the black hole event horizon\cite{Hills75} and the disruption results in a luminous flare\cite{vanVelzen11, Gezari12, Arcavi14, Holoien14ae}. 
Here we report observations over a period of 10 months of a transient, hitherto interpreted\cite{Dong16} as a superluminous supernova\cite{Quimby11}. 
Our data show that the transient rebrightened substantially  in the ultraviolet and that the spectrum went through three different spectroscopic phases without ever becoming nebular.
Our observations are more consistent with a tidal disruption event 
than a superluminous supernova 
because of the temperature evolution\cite{Holoien14ae}, the presence of highly ionised CNO gas in the line of sight\cite{Cenko16} and our improved localisation of the transient in the nucleus of a passive galaxy, where the presence of massive stars is highly unlikely\cite{Lunnan14,Leloudas15}.
While the supermassive black hole has a mass $> 10^8 M_\odot$ \cite{ReinesVolonteri15,McConnellMa13}, a star with the same mass as the Sun could be disrupted outside the event horizon if the black hole were spinning rapidly\cite{Kesden12a}.   
The rapid spin and high black hole mass can explain the high luminosity of this event.
\end{abstract}

\begin{figure}
\centering
\includegraphics[width=16cm]{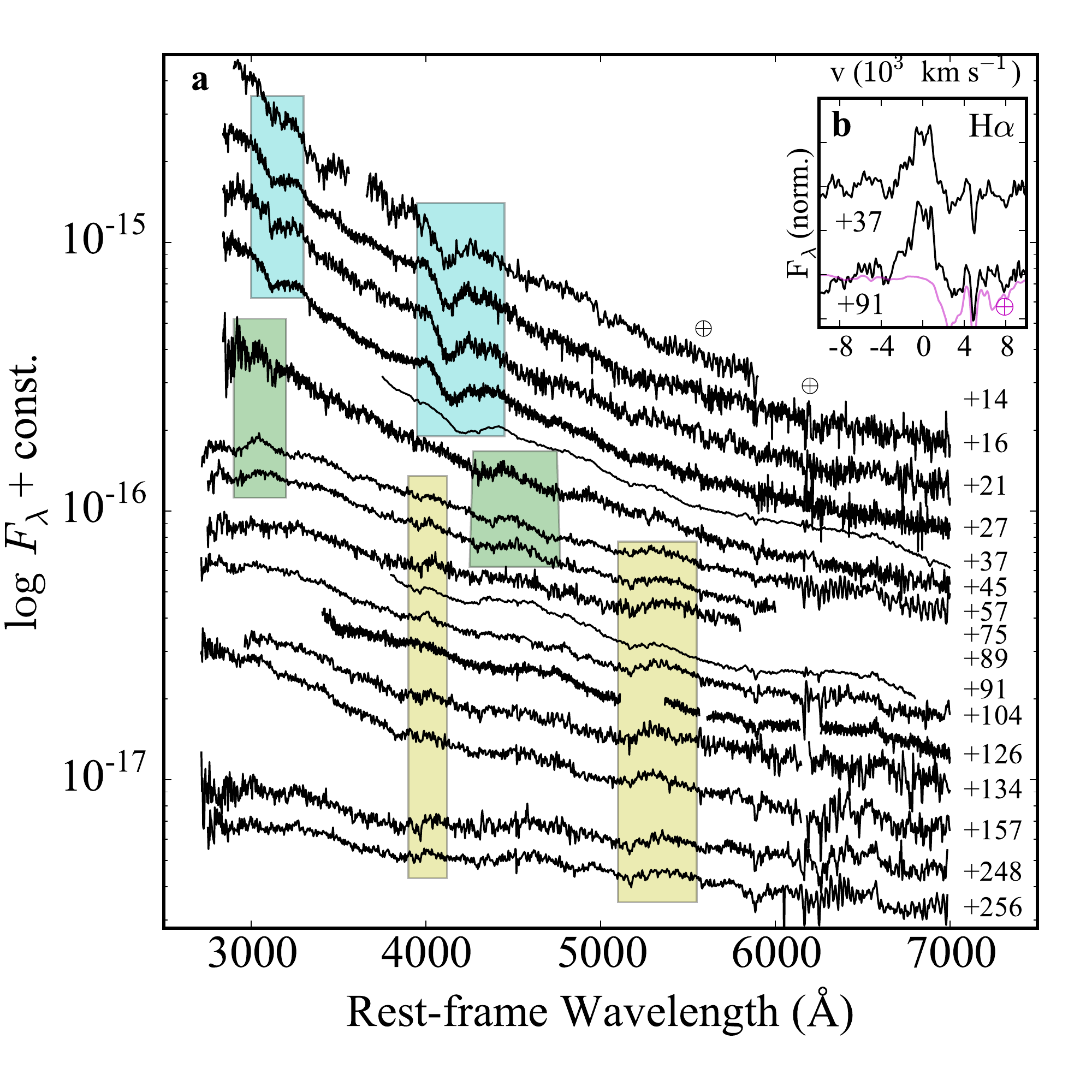}
\caption{\boldmath$\vert$\unboldmath \hspace{1em} \textbf{Spectral sequence of ASASSN-15lh showing three spectroscopic phases. a,} 
The main spectral features during the different  phases are highlighted with different colours.
The two most recent spectra appear redder due to the increased host contamination.
Rest-frame phases are indicated, the spectra have been offset for clarity and the Earth symbol marks the strongest telluric features.
\textbf{b,} Detection of H$\alpha$ (FWHM $\sim 2,500$ km s$^{-1}$) in a telluric-free region of our best spectra.
The magenta line is a telluric spectrum. 
\label{fig:specseq}
}
\end{figure}

ASASSN-15lh was discovered by the All-Sky Automated Survey for SuperNovae (ASAS-SN) on 14 June 2015 at a redshift of $z=0.2326$.  
Its light curve peaked at $V \sim 17$ mag implying an absolute magnitude of $M=-23.5$ mag, more than twice as luminous as any known supernova (SN)\cite{Dong16}. 
Our long-term spectroscopic follow-up reveals that ASASSN-15lh  went through three different spectroscopic phases (Figure~\ref{fig:specseq}). 
During the first phase\cite{Dong16}, the spectra were dominated by two broad absorption features. While these features appear similar to those observed in superluminous supernovae (SLSNe; Supplementary Figure~\ref{fig:specComp}), their physical origin is different. The features in SLSNe are due to O~\textsc{ii}\cite{Quimby11,Mazzali16}, but this would produce an additional strong feature at  $\sim$4,400 \AA\ (Supplementary Figure~\ref{fig:SynowVels}) .
The feature at $\sim$4,100 \AA\ cannot  be easily identified in the tidal disruption event (TDE) framework either. 
Two possibilities are that it could be due to absorption of Mg \textsc{ii} or high-velocity He \textsc{ii}\cite{StrubbeQuataert11}.
After the initial broad absorption features disappeared, the spectra of ASASSN-15lh were dominated by two emission features.
A possible identification for these features is He \textsc{ii} $\lambda\lambda 3,202$ and $4,686$ \AA, which are both consistently blueshifted by $\sim$15,000 km s$^{-1}$
(Supplementary Figure~\ref{fig:HeII}). 
He \textsc{ii} emission is commonly seen in optically discovered TDEs\cite{Gezari12,Arcavi14} at different blueshifts, albeit typically at lower velocities, but it has not been seen in H-poor SLSNe.
These features disappeared after day $+75$ (measured in rest frame from peak) and the later spectra were mostly featureless, with the exception of two emission features at $\sim$4,000 and 5,200 \AA. The spectra remained much bluer than those of SLSNe\cite{Pastorello10gx} for many months after the peak and never revealed nebular features, even up to day $+$256. 

A UV spectrum obtained with the  \textit{HST} on day $+168$ does not show any broad features\cite{Brown2016}. 
At the redshift of the host, we identified weak Ly-$\alpha$ absorption and disproportionally strong high-ionisation lines (N~\textsc{v}, O~\textsc{vi}, C~\textsc{iv}). 
Combined with the weakness (or absence) of common\cite{AdUP2012} low-ionisation absorption lines (Fe \textsc{ii}, Si \textsc{ii}, Mg \textsc{ii}), this 
aspect 
seems to be similar to the spectrum of ASASN-14li, the only available UV spectrum of a TDE\cite{Cenko16}  (see Methods and Supplementary Figure~\ref{fig:COS}). 
The highly-ionised gas appears at slightly different velocities, suggesting that it could be due to material in the vicinity of the TDE and ionised by its radiation. 
In contrast to the case of ASASSN-14li\cite{Cenko16}, we do not observe any broad features in
the UV range, but the optical spectrum is also mostly featureless
at these phases.

In addition, we detect the presence of hydrogen in ASASSN-15lh. A weak H$\alpha$ emission line is unambiguously detected in our highest signal-to-noise spectra (Figure~\ref{fig:specseq}) and in more spectra at lower significance (Supplementary Figure~\ref{fig:HeII}). 
Its presence cannot be excluded in any spectrum and its strength (equivalent width $\sim$4--8 \AA) is invariable, within the present errors. 
The velocity of the H$\alpha$ line (full-width at half-maximum (FWHM) $\sim 2,500$ km s$^{-1}$) is different than those of other features, implying that it  is formed in a different emitting region.

\begin{figure}
\centering
\includegraphics[width=16cm]{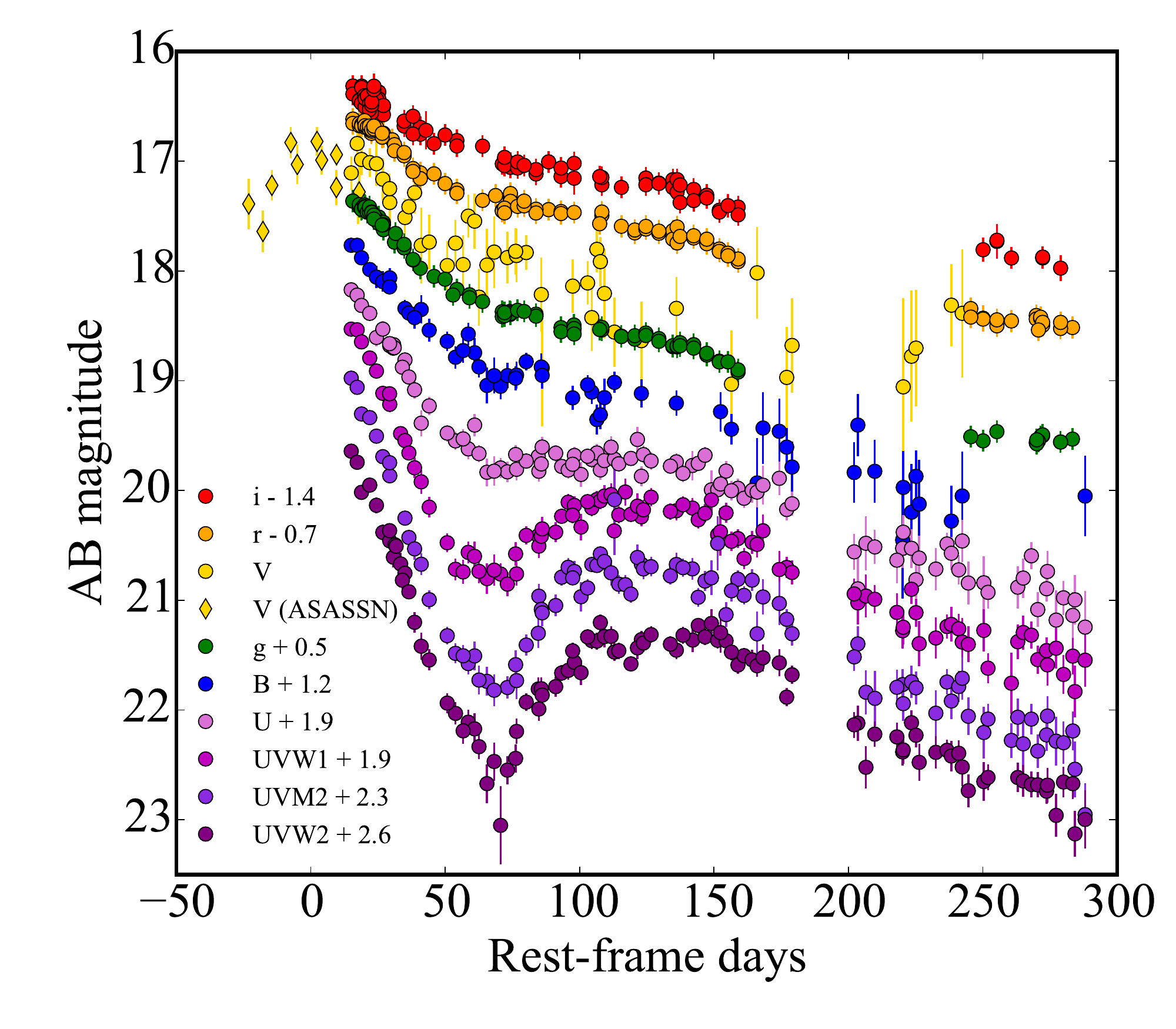}
\caption{\boldmath$\vert$\unboldmath \hspace{1em}  \textbf{The light curve evolution of ASASSN-15lh in the rest frame.} The data are from LCOGT ($gri$) and \textit{Swift} (other filters), supplemented by the ASASSN $V$-band data\cite{Dong16}. We have adopted a peak time at 5 June 2015 (MJD 57178.5)\cite{Dong16}. The light curves are shifted for clarity as indicated in the legend. 
Error bars represent 1$\sigma$ uncertainties.
The optical bands show a monotonic decline, but the UV bands show a rebrightening after 60 rest-frame days. 
A significant secondary dip is also observed in the bluest bands around day +120.
The photometry has been corrected for foreground extinction and the host contribution has been removed (see Methods).
\label{fig:LC}
}
\end{figure}

\begin{figure}
\centering
\includegraphics[width=16cm]{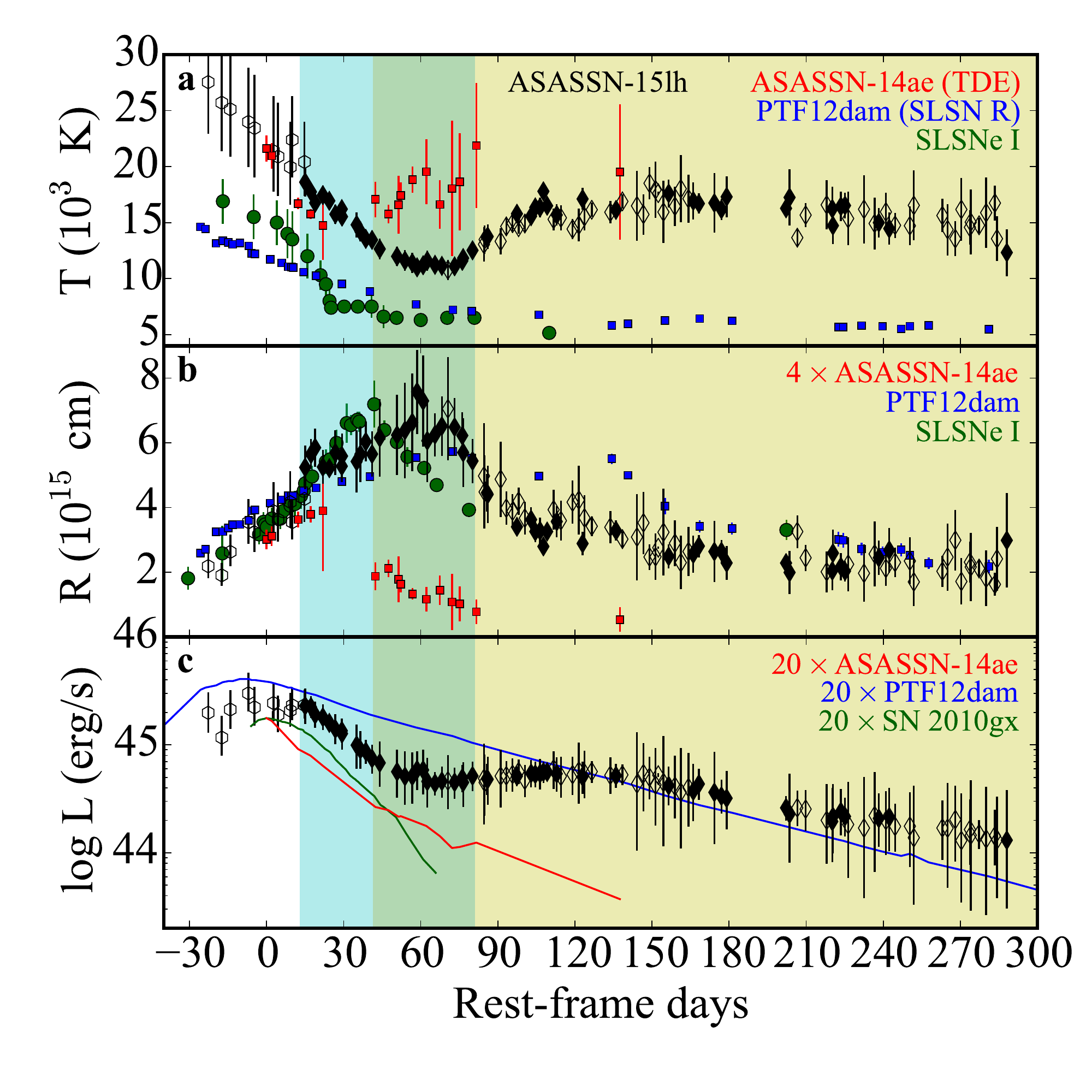}
\caption{\boldmath$\vert$\unboldmath \hspace{1em} \textbf{The evolution of the temperature, radius and luminosity of ASASSN-15lh, compared with TDEs\cite{Holoien14ae} and SLSNe\cite{Pastorello10gx,Inserra2013,Nichollpessto,Chen12dam}}.
The coloured areas correspond to the different spectroscopic phases in Figure~\ref{fig:specseq}.
For ASASSN-15lh (data in black), open symbols show fits based on fewer than five filters.
In particular, open hexagon symbols show early data, based only on the $V$ band and derived with a temperature prior\cite{Dong16}.
Error bars represent 1$\sigma$ uncertainties.
The curves are shown with respect to peak time and some comparison objects have been scaled  as indicated in the legend.
\textbf{a,b,} The evolution of the temperature \textbf{(a)} and radius \textbf{(b)}  of ASASSN-15lh are qualitatively similar to those of the TDE ASASSN-14ae, although this is happening in longer timescales and larger radii.
All types of SLSNe cool down with time. 
\textbf{c,} The bolometric luminosity  of ASASSN-15lh shows an extended plateau between 70 and 160 days.
\label{fig:BBevol}}
\end{figure}

The light curve evolution of ASASSN-15lh is  shown in Figure \ref{fig:LC}. After the initial peak and decline, around 10 September  (day $+60$), the UV started rebrightening, an effect that was more prominent in the far-UV bands\cite{2015ATel.8086....1B,GodoyRivera,Brown2016}.
The dense photometric follow-up with the Swift Gamma-Ray Burst Mission (\textit{Swift}) and the Las Cumbres
Observatory Global Telescope Network (LCOGT) revealed that ASASSN-15lh reached a secondary UV maximum at around $+110$ days, followed by another decline.
Interestingly, after day $+100$, the colours of ASASSN-15lh  remained almost constant  for over 120 rest-frame days (Supplementary Figure \ref{fig:colors}).
By fitting a black body to the multi-wavelength photometry of ASASSN-15lh, 
we are able to estimate the temperature evolution,  black-body radius and bolometric luminosity (Figure \ref{fig:BBevol}). 
While the UV rebrightens and the spectrum changes, the blackbody radius decreases and the temperature increases again, stabilising at $\sim 16,000$ K.
This is neither expected from a  
SN photosphere, nor observed for SLSNe\cite{Inserra2013,Nichollpessto,Chen12dam}.
However, the TDE candidate ASASSN-14ae\cite{Holoien14ae} showed a very similar temperature evolution to ASASSN-15lh.
Even if this happened at shorter timescales, the qualitative similarity between the evolution of the two events suggests that they might be due to the same mechanism.
On the other hand, the radius of ASASSN-15lh is larger by a factor of about seven, and ASASSN-14ae has a much stronger H$\alpha$ line (Supplementary Figure~\ref{fig:specComp}).

By integrating the bolometric luminosity, we estimate that ASASSN-15lh had radiated a total of $1.88 \pm 0.19 \times 10^{52}$ erg (depending on the assumed bolometric correction) up to 25 May 2016  (day +288). 
Including kinetic energy, which can reach an additional $10^{52}$ erg for SLSNe\cite{Mazzali16,Howell13}, 
the total energy budget approaches the theoretical limit of that which SN explosions models can accommodate\cite{Metzger15,Sukhbold16}.
It is possible that UV rebrightening could occur in a SLSN, due to either strong circumstellar interaction or the ionization breakout powered by a central magnetar. However, the observed H$\alpha$ line is much weaker than those in SLSNe that have shown signs of late interaction\cite{Yan13ehe} (Supplementary Figure~\ref{fig:specComp}), and there are no features indicative of interaction in the UV spectrum.
In addition, predictions for ionisation breakout suggest that the spectrum should turn nebular\cite{Metzger15}, although this might apply better to an X-ray rather than a UV breakout.
Nevertheless, no detailed model has yet been calculated that can naturally explain the entire spectroscopic and photometric properties observed, either in the SN or in the TDE scenario. 
A single epoch of imaging polarimetry with \textit{HST} shows low levels of polarisation\cite{Brown2016}, similar to that obtained for a SLSN\cite{LeloudasLSQ14mo}, and suggesting an only mildly asymmetric geometry (in projection). Polarisation measurements and predictions for optical TDEs are still lacking.

Strong evidence for ASASSN-15lh being a TDE comes from its environment.
H-poor SLSNe are found in blue, metal-poor dwarf galaxies 
with average masses of $\log _{10}M_{\star}=8.24 \pm 0.58$  $M_\odot$, and none have yet been  found to exceed $9.60$ at $z<1$ \cite{Lunnan14,Leloudas15}. 
These galaxies typically have strong emission lines, pointing to active ongoing star formation and young progenitor ages that do not exceed a few Myr\cite{Leloudas15}. 
In contrast, the host of ASASSN-15lh is a massive and passive red galaxy. 
By fitting the available photometry (see Methods), we estimate that the mass of the host is $\log _{10}M_{\star}=10.95^{+0.15}_{-0.11}$ $M_\odot$, with a dominant stellar population of age $3.9^{+3.2}_{-1.3}$ Gyr. The spectral energy distribution (SED) fit provides a star formation rate (SFR) of $0.05^{+0.15}_{-0.05}$ $M_\odot$ yr$^{-1}$, consistent with the improved limit on SFR of $< 0.02$ $M_\odot$ yr$^{-1}$ that we obtained from our highest signal-to-noise spectrum.
The derived specific star formation rate of $\log{\rm{sSFR}} < -12.5$ yr$^{-1}$ is thus three orders of magnitude lower than in any H-poor SLSN host (Supplementary Figure~\ref{fig:hostsComp}).
Furthermore, the transient is positionally coincident with the nucleus of its host.
By aligning postdiscovery \textit{HST} Advanced Camera for Surveys (ACS) images with a prediscovery image taken with the Cerro Tololo Inter-American Observatory
4 m Dark Energy Camera (Supplementary Figure \ref{fig:astrom}), we were able to improve the positional accuracy\cite{Dong16} of ASASSN-15lh by a factor of about four, corresponding to a projected nuclear offset of $131 \pm 192$ pc.

\begin{figure}
\centering
\includegraphics[width=16cm]{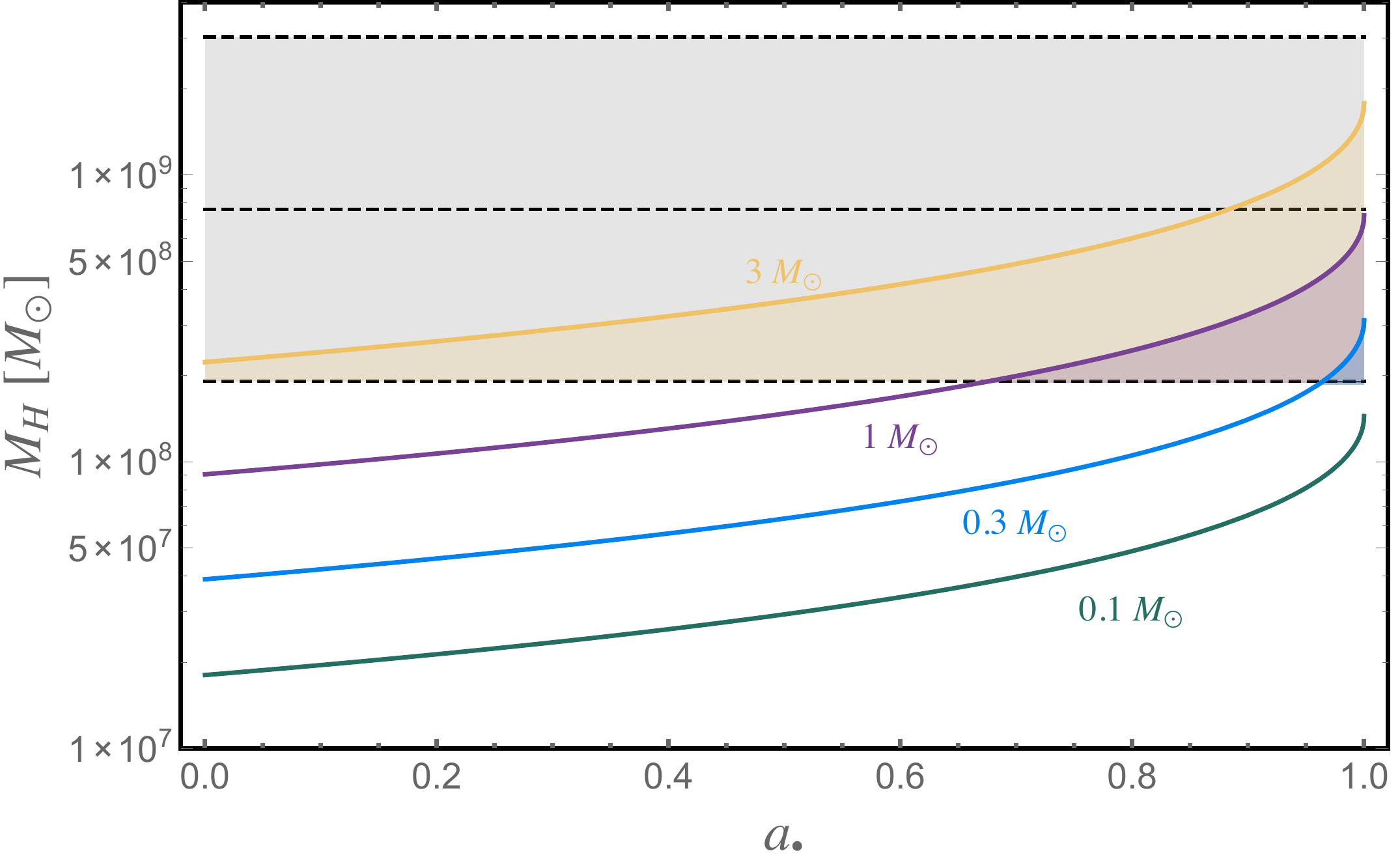}
\caption{\boldmath$\vert$\unboldmath \hspace{1em}  \textbf{
The Hills mass $M_{\rm H}$  as a function of SMBH spin $a_\bullet$ for main sequence stars of different masses. }
The SMBH mass estimate\cite{ReinesVolonteri15} for the host of ASASSN-15lh and the corresponding $1\sigma$ uncertainty region are shown as dashed lines and grey shading.  
The plot shows that a star of mass $0.1 M_\odot$ cannot be disrupted by the SMBH, as the Hills mass is always below the mass of the SMBH. Stars of mass $0.3 M_\odot$ and $1 M_\odot$ can be disrupted by rotating Kerr black holes of different spins. 
Only stars of mass $3 M_\odot$  lie in the allowed region for TDEs from a non-rotating Schwarzschild black hole, but TDEs from such stars are unlikely\cite{StoneMetzger16,Kochanek2016}.
\label{fig:HillsMass}}
\end{figure}

It has been argued that the large host galaxy mass may imply the presence of a supermassive black hole (SMBH) that is too large to disrupt stars outside its event horizon\cite{Dong16}.  
Since the tidal radius scales as $R_{\rm t} \propto M_\bullet^{1/3}$ while the gravitational radius scales as $R_{\rm g} \propto M_\bullet$, stars can only be disrupted outside the horizon of a SMBH if the black hole is below a certain size, the Hills mass\cite{Hills75} $M_{\rm H}$.  Larger SMBHs swallow stars whole.  For a non-spinning Schwarzschild SMBH, the Hills mass is $M_{\rm H} \approx 9 \times 10^7 M_\odot r_\star^{3/2}m_\star^{-1/2}$, where $m_\star \equiv M_\star / M_\odot$ and $r_\star \equiv R_\star / R_\odot$ (see Methods).  
Using an empirical relationship between SMBH mass and total stellar mass for elliptical and
spiral/lenticular galaxies with classical bulges\cite{ReinesVolonteri15}, we find $\log _{10}M_\bullet =8.88\pm 0.60$  $M_\odot$, far above the Schwarzschild Hills mass for solar-mass stars.  Using an $M_\bullet - L$ relation for early-type galaxies\cite{McConnellMa13}, we obtain  $\log _{10}M_\bullet =8.50\pm 0.52$ $M_\odot$.  However, $M_{\rm H}$ increases by almost an order of magnitude for rapidly spinning Kerr SMBHs and favorable orbital orientations\cite{Kesden12a}.  For an optimal (prograde equatorial) orbit and our range of SMBH mass estimates, we find that a solar-mass star can be  disrupted by a SMBH with dimensionless spin parameter $a_\bullet = 0.68$  if $\log _{10}M_\bullet = 8.28$ $M_\odot$,
and by a SMBH with  dimensionless spin parameter $a_\bullet = 1$ if $\log _{10}M_\bullet = 8.86$ $M_\odot$.
We show the exact relativistic $M_{\rm H}(a_\bullet)$ in Figure~\ref{fig:HillsMass}. 
For stars less massive than the Sun, the spin is constrained to even higher values.
ASASSN-15lh could be compatible with a TDE by a Schwarzschild SMBH provided $M_\star \gtrsim 3 M_\odot$.
However, the typical tidally-disrupted star comes from the lower end of the stellar mass function, and this hypothesis is further challenged by the old age of the galaxy's stellar population\cite{StoneMetzger16,Kochanek2016}.  
Observations of active galactic nuclei suggest that rapid SMBH spins are common\cite{ReynoldsSreview}.
 We demonstrated
here that TDEs present a method to probe the SMBH  spins of quiescent galaxies.
Given the inferred rapid spin of the SMBH, the fact that we did not detect a jet at radio wavelengths implies that black hole spin alone is not sufficient to launch powerful jets (see Methods).

The luminosity and energetics of ASASSN-15lh are also explained by a particularly  massive SMBH. 
The expected radiative efficiency of accretion increases from $\eta \approx 0.05$ (for a Schwarzschild SMBH) to $\eta \approx 0.42$ for a rapidly spinning SMBH disrupting stars on prograde, near-equatorial orbits.  
A particularly massive SMBH is further biased towards disrupting the most massive stars near the main sequence turn-off mass, increasing accretion rates and total energy release.  
Finally, it is known that most TDEs only release a small fraction of $\eta M_\star c^2$ in accretion power (the so-called ``missing energy problem'' \cite{StoneMetzger16, Piran15}).
One compelling explanation for this is that circularization of debris and formation of the accretion disk is mediated by relativistic apsidal precession, and that the majority of TDEs circularize inefficiently due to weak apsidal precession\cite{Dai15}.  
For a particularly massive fast-spinning SMBH, efficient circularization is favoured because $R_{\rm t} \sim R_{\rm g}$, ensuring large per-orbit precession.  
The peculiar light curve may also be a natural consequence of tidal disruption by extremely massive black holes. 
In Methods we combine two competing models (the ``circularization''\cite{Piran15} and the ``accretion/reprocessing'' \cite{Guillochon14} paradigms) for the optical emission in TDEs and show that the most massive SMBHs produce an unusual hierarchy of tidal disruption timescales.  Since the viscous time in the accretion disk is much longer than the debris fallback time when $M_\bullet$ is greater than a few times $10^7 M_\odot$, TDEs around the most massive SMBHs can display an early peak in the light curve from circularization luminosity, and a second peak from reprocessed accretion luminosity.



 \clearpage
 
 \begin{addendum}

\item[Correspondence] Correspondence and requests for materials
should be addressed to Giorgos Leloudas.~(email: giorgos@dark-cosmology.dk).

\item 
We acknowledge support from the European Union FP7 programme through the following ERC grants: 320360 (MF, HC), 647208 (PGJ), 291222 (SJS), 615929 (MS).
We also acknowledge: Einstein Postdoctoral Fellowship PF5-160145 (NCS), Hubble Postdoctoral Fellowship HST-HF2-51350 (SvV), STFC grants ST/I001123/1 ST/L000709/1 (SJS) and ST/L000679/1 (MS), Australian Research Council Future Fellowship FT140101082 (JCAMJ), a Royal Society University Research Fellowship (JRM), a Sofja Kovalevskaja Award to P. Schady (TKr, TWC), a Ram\'{o}n y Cajal fellowship and the Spanish research project AYA 2014-58381 (AdUP), CONICYT-Chile FONDECYT grants 3130488 (SK), 3140534 (SS), 3140563 (HK), 3150238 (CRC), a PRIN-INAF 2014 project (NER), support from IDA (DM), an Ernest Rutherford Fellowship (KM), CAASTRO project number CE110001020 (BET), NSF Grant AST 11-09881 and NASA Grant HST-AR-13726.02 (JCW).
This work makes use of observations from the Las Cumbres Observatory Global Telescope Network and is based upon work supported by the NSF Grant No. 1313484.
The Australia Telescope Compact Array is part of the Australia Telescope National Facility which is funded by the Australian Government for operation as a National Facility managed by CSIRO.
Based partially on observations collected as part of PESSTO (the Public ESO Spectroscopic Survey for Transient Objects Survey) under ESO programs188.D-3003, 191.D-0935 and on observations made with ESO Telescopes at the La Silla Paranal Observatory under programme ID 095.D-0633.
We thank Massimo Della Valle for comments.

 \item[Author Contributions] 
 GL coordinated the PESSTO observations, was PI of the FORS2 program, analysed the data and wrote the paper. 
 MF provided the astrometric localisation and reduced the PESSTO spectra. 
 NCS calculated the relation between the BH spin and the Hills mass and edited the manuscript. 
 SvV performed the \textit{Swift} photometry. 
 PGJ analysed XMM data and helped coordinating the project. 
 IA is the PESSTO PI for TDEs and provided LCOGT data. 
 CF made the LCOGT photometry. 
 JRM reduced the FORS2 spectra. 
 SJS is the PI of PESSTO and helped coordinating the project. 
 TKr provided the SED fit of the host galaxy.
 JCAMJ provided the radio observations.
 PMV helped with the analysis of the spectra.
 GL, MF, NCS, SvV, PGJ, IA, SJS, JCAMJ, AG-Y and PAM contributed to the discussions.
 AdUP and ADC worked on the UV spectrum.
 DAH is the PI of the LCOGT observations.
 CI and OY are PESSTO builders and helped with the analysis.
 FP, DM, JS and JCW provided FORS2 data and analysis.
 MC and BET provided the WIFES spectra.
 SS and SK provided the Magellan spectrum.
 GH, CMcC and SV obtained and reduced LCOGT data. 
 EK, KM, KWS, MS, and DRY are PESSTO builders and CA, JH, SM, TWC, TKa, SP, CRC, HK, MN, JL, NER, HC, IB, JJ, MRM contributed with PESSTO observations or data   reductions.
Many authors provided comments on the manuscript.

 \item[Competing Interests] The authors declare that they have no
competing financial interests.

\end{addendum}


\clearpage

\newpage

\newpage

\newpage

\clearpage

\begin{methods}

We describe here the data, methods and theoretical calculations used. We provide
details on the localization of ASASSN-15lh and on the host galaxy properties.
We present the different data used, and describe the reductions and comment
on the implications. Finally, we demonstrate that a TDE can easily accommodate
the luminosity and energetics of ASASSN-15lh, we show that combining two
luminosity mechanisms for TDEs can result in light curves with two different
timescales and we present our derivation of the relativistic Hills mass as a function of the SMBH spin. 
We assume a \textit{Planck} cosmology\cite{2014A&A...571A..16P}.

\subsection{Astrometric localisation of ASASSN-15lh}

\textit{HST} observations of ASASSN-15lh were obtained under programme 14348 (PI: Yang) with ACS and a broad-band polarimetry filter (POL0V) and the F606W filter.
All analyses
were performed on the reduced drizzled image 
obtained from the MAST archive. The image was taken on 2015 August 3, and the total exposure time for this
frame was 366~s. The DECam image was a 90 s image taken on 2014 October 22 using an
$r$-band filter. The DECam data were reduced using the DES Data Management Pipeline and are
available online\cite{2015ATel.7843....1M}. The measured FWHM was 0.8$''$.

Eight sources common to both the DECam and ACS images were used to derive a geometric
transformation (allowing for rotation, translation and a single scaling
factor) between the two frames. Of these sources, five were unresolved in
the ACS image and the remaining three had a FWHM of less than about twice that
of the point sources. 
The root mean square error error in the transformation was 0.19 DECam pixels, or 52 mas.
To measure the position of the host galaxy in the DECam frame, we fitted a model galaxy
profile using the {\sc galfit} code; uncertainties in the galaxy centroid  were estimated to be
only a few $\times$0.01 pixels from Monte Carlo tests.
The position of ASASSN-15lh on the ACS image yields an accuracy of 0.05 pixels (or 3 mas) using 
 three different centering algorithms within the {\sc iraf} {\sc phot} package.

\subsection{The host galaxy}

To derive the physical properties of the host galaxy of ASASSN-15lh through modeling 
of its spectral energy distribution (SED), we used $grizY$ \cite{2015ATel.7843....1M}, 
$J$ and $K_\mathrm{s}$ \cite{2015ATel.7776....1P}, as well as $3.4\,\mu\mathrm{m}$ and 
$4.6\,\mu\mathrm{m}$ \textit{WISE} photometry. We also performed aperture photometry on  
\textit{GALEX} images 
yielding no significant 
detections.

We fitted the Galactic extinction-corrected\cite{2011ApJ...737..103S} photometry of the host with stellar population synthesis models\cite{2003MNRAS.344.1000B} using the code \textit{Le Phare}\cite{1999MNRAS.310..540A, 2006A&A...457..841I}. 
Our galaxy templates were based on a Chabrier initial mass function\cite{2003PASP..115..763C}, 
and spanned different stellar metallicities, 
e-folding timescales $\tau$  (0.1 to 30 Gyr), stellar population ages (0.01 to 10 Gyr) and dust attenuations\cite{2000ApJ...533..682C}. 
The galaxy stellar mass and SFR are  $\log _{10}M_{\star}=10.95^{+0.15}_{-0.11}$ M$_\odot$ and $0.05^{+0.15}_{-0.05}$ M$_\odot$ yr$^{-1}$, respectively. 
Physical parameters are given as the median of the probability distribution 
of all templates, with error-bars containing the $1\sigma$ probability interval. 
The best fit model is shown in Supplementary Figure \ref{fig:SEDfit} and it has a low $E(B-V)_\mathrm{host} = 0.02$ mag. 
Throughout the paper, we assume that the extinction at the host is negligible.

We also constrain the recent star formation in the host by placing limits on the flux of [O~\textsc{ii}] and (narrow) H$\alpha$.
Using the FORS2 spectra (signal-to-noise ratio $> 200$) we obtain flux limits  of $<1.8 \times 10^{-16}$ erg s$^{-1}$ cm$^{-2}$  for [O \textsc{ii}] and $<2.9 \times 10^{-17}$ erg s$^{-1}$ cm$^{-2}$ for H$\alpha$ (2$\sigma$). These limits constrain the SFR to $<0.25$ and $<0.02$ M$_\odot$ yr$^{-1}$ respectively\cite{1998ARA&A..36..189K,2003PASP..115..763C}, an improvement by a factor of ten over previous estimates\cite{Dong16}.

\subsection{\textit{HST} UV spectroscopy}

A UV spectrum of ASASSN-15lh was obtained on day $+168$ with \textit{HST} under program 14450 (PI: Brown).
We downloaded the reduced COS and STIS spectra from the MAST archive. 
The spectrum does not display any broad emission or absorption features (at similar phases the optical spectrum is also mostly featureless; Figure~\ref{fig:specseq}).
We identified both geocoronal and absorption lines at $z = 0$ and a number of narrow (FWHM $\sim 200-400$ km s$^{-1}$) absorption features at the redshift of ASASSN-15lh. 
Supplementary Figure~\ref{fig:COS} shows the COS spectrum (the STIS spectrum is more noisy and less interesting).
Supplementary Table~\ref{tab:abslines} contains the EWs and kinematical offsets (measured relative to Ly-$\alpha$) for lines that were detected, as well as selected limits.

Low-ionisation features, such as Fe \textsc{ii} and Si \textsc{ii}, which are common in the star-forming sightlines  of SLSNe\cite{Vreeswijk14} or GRBs\cite{Christensen2011,AdUP2012} are weak or absent in ASASSN-15lh.  
In contrast, absorption from high ionisation lines from N \textsc{v} and O \textsc{vi} is remarkably strong, especially relatively to the (weak) Ly-$\alpha$. 
In particular, the ratio of N \textsc{v} to Ly-$\alpha$ is $\sim$4, while it is $>$1,000 lower in GRBs. 
The same is true for column densities: by Voigt profile modelling, we derived $\rm{N(H~\textsc{i})} = 14.73 \pm 0.12$, $\rm{N(O~\textsc{vi})} = 15.58 \pm 0.03$, $\rm{N(N~\textsc{v})} =  15.42 \pm 0.06$ and $\rm{N(C~\textsc{iv})} = 14.60 \pm  0.27$, resulting in ratios that are highly unusual for GRB or Quasar DLAs\cite{2007A&A...465..171F,2008A&A...491..189F},
even if those values are lower limits due to saturation.
The ratio of N \textsc{v} to Ly-$\alpha$ was also observed to be of the order of unity in the spectrum of ASASSN-14li, the only UV spectrum of a TDE\cite{Cenko16}. 
By complementing our measurements with those from optical spectra\cite{Dong16}, we found that the Mg \textsc{ii} absorption is weak, below the value  for GRBs\cite{Christensen2011,AdUP2012} and  SLSNe\cite{Vreeswijk14}.

Therefore we suggest that the absorbing gas can be separated into two components: (1) a tenuous mildly-ionised medium (Ly-$\alpha$, Mg \textsc{ii}, Si \textsc{iii} at velocities from 0 to +44 km s$^{-1}$), which is very unusual given the lack of Fe \textsc{ii} and Si \textsc{ii}, and (2) a highly-ionised medium (N \textsc{v}, O \textsc{vi} and C \textsc{iv}) at negative velocity offsets $-$80 to $-$120 km s$^{-1}$. We suggest that the latter is consistent with material from a disrupted low-mass star\cite{Kochanek2015,Cenko16} and that it is ionised by the TDE. Despite the absence of broad features, the  phenomenological similarity with the UV spectrum of ASASSN-14li strongly favours a TDE origin for ASASSN-15lh. A highly-ionised outflow was also detected for ASASSN-14li in X-rays\cite{Miller2015}.

\subsection{Optical spectroscopy}

Spectra were obtained with the instruments and set-ups listed in Supplementary Table~\ref{tab:speclog}. The FLOYDS, WiFeS and EFOSC2 data were reduced using dedicated instrument pipelines \cite{dopita07, dopita10, SmarttPessto}. The VLT+FORS2 and Magellan+IMACS spectra were reduced in the standard fashion using IRAF. The FORS2 spectra were obtained in spectropolarimetric mode, but the ordinary and extraordinary rays were combined to produce an intensity spectrum.

\subsection{Imaging and removal of the host contribution}
The LCOGT $gri$ images were pre-processed using the Observatory  Reduction  and  Acquisition  Control  Data  Reduction  pipeline\cite{2015A&C.....9...40J}. To remove the host contribution, we performed image subtraction using the pre-discovery DECam $gri$ images as templates.
The \textit{Swift} UVOT observations were reduced following the standard procedures and software (\textsc{uvotsource}). To extract the photometry, we used a 4$''$ aperture and a curve of growth aperture correction. 
For the \textit{Swift} filters we did not have pre-discovery observations, and hence used the model galaxy spectrum from the SED fit (Supplementary Figure~\ref{fig:SEDfit}) to generate synthetic magnitudes at these wavelengths. For the $B$ and $V$ filters, where host contamination is a concern, we estimated the host uncertainty to be $< 0.05$ mag, increasing to $0.1$ mag for the $U$ band. The host uncertainties in the UV filters are more significant, but at these wavelengths the host is many orders of magnitude below the transient luminosity. In the AB system we obtained host magnitudes of 
$ V 	=  18.98,  B 	= 	20.43, U 	=	21.95, UVW1 =	23.21, UVM2 =	23.65$ and $UVW2 =	23.62$~mag, which we subtracted from the \textit{Swift} measurements to obtain the transient photometry.
Two UVOT  filters suffer from a red leak but this does not affect blue sources as ASASSN-15lh to the same degree as e.g. SNe Ia\cite{Brown2010}.
Based on synthetic photometry of black-body spectra, we estimated that the photometry is affected by $< 2\%$ for a black body with $T =15,000$ K. As this precision is significantly lower than our photometric accuracy, we did not attempt to correct for this effect. 
Another study\cite{Brown2016} has found higher values -- but still low -- for this maximum contamination.

\subsection{Radio and X-ray observations}

We observed ASASSN-15lh from 05:00 to 14:00\,UT on 9 December 2015, using the Australia Telescope Compact Array in 750C configuration, under project code CX340.  We observed in two frequency bands of width 2.048\,GHz, centered at 5.5 and 9.0\,GHz.  We used B1934-638 as both our flux and bandpass calibrator, and B2205-636 as our phase calibrator.  We reduced the data following standard procedures in Miriad\cite{Sault1995}, and carried out the imaging and self-calibration using the Common Astronomy Software Application\cite{McMullin2007}.

The field was dominated by PKS\,J2203-6130, a 9-mJy source (prior to primary beam correction) located 15\,arcmin\ away from the target.  We performed self-calibration, initially in phase only (down to a timescale of 2\,min), and eventually in amplitude and phase, on a timescale of 10\,min.  Given the non-uniform {\it uv}-coverage, we tested a variety of image weighting schemes, and found the optimum to be a Briggs robust weighting of 0.5.  ASASSN-15lh was not detected down to $3\sigma$ upper limits of 25 and 23\,$\mu$Jy\,beam$^{-1}$ at 5.5 and 9.0\,GHz, respectively, consistent with a reported upper limit from three weeks earlier\cite{Kool2015}.  Stacking our two frequency bands gave us a slightly deeper $3\sigma$ radio upper limit of 17\,$\mu$Jy\,beam$^{-1}$.

The XMM--{\it Newton} satellite observed ASASSN--15lh as part of a Directors Discretionary Time proposal on 18 November  2015.  The on--source time is 11.9 ks, and after filtering epochs of high background, 9 ks of MOS2, and 4 ks of PN data can be used.  Given the lower background and the longer net exposure, we used the MOS2 detector for estimating the upper limit on the source flux. In an aperture of radius 32$"$ centred on ASASSN-15lh we derived a 95\% confidence upper limit\cite{1991ApJ...374..344K,1984NIMPA.228..120H} of 11 source counts in the 0.15--1 keV band. To convert this to a limit on the flux, we use a temperature of 70 eV found for ASASSN--14li as input, which taking into account that the M$_{\bullet}$ in ASASSN--15lh is close to 8$\times 10^{8}$ M$_\odot$, and that of ASASSN--14li is closer to 2$\times 10^6$ M$_\odot$, implies a blackbody temperature of 18 eV (as $T\propto M^{-0.25}$). With this and N$_H$=3$\times$20 cm$^{-2}$ W3PIMMS provides a 95\% upper limit to the 0.3--1 keV X--ray flux of 2$\times 10^{-16}$ erg cm$^{-2}$ s$^{-1}$, yielding an upper limit to the source luminosity  of 3$\times 10^{40}$ erg s$^{-1}$.
This limit depends strongly on the chosen energy band. The 0.3--1 keV band was chosen to allow comparison with ASASSN-14li\cite{vanVelzen14li}.

If TDEs do in fact all produce radio jets, as suggested by the recent detection of ASASSN-14li\cite{vanVelzen14li}, then applying the same model with appropriate scalings would predict a radio flux of 10\,$\mu$Jy at the time of the observations.  The fact that
we did not detect a radio jet therefore remains consistent with a TDE origin for
ASASSN-15lh, and implies that the jet power of ASASSN-15lh is $\nu L_{\nu} \lesssim 10^{38}$ erg s$^{-1}$, which is at least 2-3 orders of magnitude lower than that of the relativistic TDE Swift J1644+57\cite{Levan11}.  
This difference was probably caused by a combination of misalignment of the jet axis and the line of sight (consistent with our non-detection of X-ray) and differences in accretion flow geometry, interstellar medium density or magnetic field strength\cite{2014MNRAS.437.2744T}. 
If black hole spin were the dominant factor in setting jet power, we would have expected ASASSN-15lh to host a jet at least comparable in strength to ASASSN-14li (as we have inferred a high black hole spin for ASASSN-15lh).  Given that our upper limit is close to the ASASSN-14li model prediction, this jet should have been detectable unless the spin of ASASSN-15lh is significantly less than we have inferred, the ISM density is extremely low, or the SMBH in ASASSN-14li is also rapidly spinning.

\subsection{Tidal Disruption Luminosities and Energetics}

In Newtonian theory, a star with mass $M_\star \equiv m_\star M_\odot$ and radius $R_\star \equiv r_\star R_\odot$ will be tidally disrupted when it approaches an SMBH of mass $M_\bullet \equiv M_8 10^8 M_\odot$ within a distance:
\begin{equation}
R_{\rm t} = R_\star \left( \frac{M_\bullet}{M_\star} \right)^{1/3}
\end{equation}
Stars passing within this tidal radius can create a luminous electromagnetic flare provided $R_{\rm t} \gtrsim R_{\rm IBCO}$, the location of the innermost bound circular orbit (IBCO) for the SMBH.  This occurs for SMBHs smaller than the Hills mass\cite{Hills75},
\begin{equation}
M_{\rm H} \equiv 9\times 10^7 M_\odot r_\star^{3/2} m_\star^{-1/2}
\end{equation}
Once the star has been disrupted, half of its gas is unbound from the SMBH; the other half remains bound with a characteristic spread in specific orbital energy \cite{Guillochon13, Stone13}
\begin{equation}
\Delta \epsilon = \frac{GM_\bullet R_\star}{R_{\rm t}^2}
\end{equation}
If we assume a top-hat distribution of debris energy, then the fallback time (in units of seconds) for the most tightly bound debris is
\begin{equation}
t_{\rm f} = 3.5 \times 10^7 ~M_8^{1/2}m_\star^{-1}r_\star^{3/2} \label{eq:tFall}
\end{equation}
which gives a peak fallback rate
\begin{equation}
\frac{\dot{M}_{\rm p}}{\dot{M}_{\rm Edd}} = 0.13 \eta_{-1} M_8^{-3/2} m_\star^2 r_\star^{-3/2} \label{eq:MDotPeak}
\end{equation}
where $\eta = 0.1 \eta_{-1}$ is the radiative efficiency of accretion used to calculate the Eddington-limited mass inflow rate $\dot{M}_{\rm Edd}$. 
The peak bolometric luminosity is 
\begin{equation}
L_{\rm p} = 1.9 \times 10^{45} \eta_{-1} M_8^{-1/2}m_\star^2 r_\star^{-3/2}~{\rm erg}~{\rm s}^{-1}
\end{equation}
which is easily compatible with our observations for $m_\star=1$ and $\eta = 0.42$, appropriate for Kerr black holes.  The radiated bolometric energy is 
\begin{equation}
E_{\rm rad} = 8.9 \times 10^{52}~ \eta_{-1} m_\star ~{\rm erg}
\end{equation}
ASASSN-15lh, although extreme as a SN \cite{Metzger15,Sukhbold16,Bersten16,Kozyreva16,Chatzopoulos16}, does not strain the theoretical luminosity or energy budgets of TDEs.

\subsection{Tidal Disruption Flare Timescales}

Existing observations do not yet distinguish between two competing theories for producing TDE optical emission.  In the accretion/reprocessing paradigm, the bolometric luminosity of a compact ($\sim 10R_{\rm g}$), efficiently circularized accretion disk is intercepted by an optically thick screen of gas at larger scales ($\sim 10^3 R_{\rm g}$).  This reprocessing layer may be a slow outflow from the disk \cite{MetzgerStone15}, or a more hydrostatic configuration \cite{Coughlin+14, Guillochon14, RothTDE}.  However, in both cases it absorbs and re-emits a large fraction of the disk bolometric luminosity at longer wavelengths.  Both cases predict a larger reprocessing layer for larger SMBHs: outflow velocities $v_{\rm w} \sim \sqrt{GM_\bullet / R_{\rm t}} \propto M_\bullet^{1/3}$, or, if we assume that a hydrostatic reprocessing layer exists at scales comparable to the semimajor axis of the most tightly bound material, its size is proportional to $M_\bullet^{2/3}$.  Alternatively, in the circularization paradigm, observed optical emission arises from shocks between debris streams\cite{Piran15}, which thermalize and radiate stream kinetic energy at the (generally large) stream self-intersection radius.

The efficiency of circularization (and therefore $\dot{M}_{\rm p}$) depends sensitively on the dimensionless orbital pericenter $R_{\rm p}/R_{\rm g}$: small decreases in $R_{\rm p}$ quickly move the self-intersection point inward, increasing circularization efficiency \cite{Hayasaki+13, Shiokawa15}.  The low luminosities seen in many optically-selected TDEs may reflect that most TDEs have sub-relativistic pericenters and circularize inefficiently, so that their peak luminosity is $\ll \eta \dot{M}_{\rm p}c^2$ \cite{Dai15}.

For an extremely massive SMBH with $M_\bullet \approx M_{\rm H}$, all TDE pericenters are highly relativistic and circularization is maximally efficient (except possibly if the SMBH spin is rapid and strongly misaligned\cite{Hayasaki+15, Guillochon15}).  Stream self-intersection points will be $\sim R_{\rm p}$, near the IBCO \cite{Dai15}, so the circularization and accretion power will be comparable.  It is therefore natural to expect that TDEs from the most massive SMBHs will be brighter and able to attain their theoretical peak  luminosity, while smaller SMBHs will often produce TDEs that (due to inefficient circularization) have luminosities well below the peak theoretical luminosity.

We propose that the two peaks in the ASASSN-15lh light curve correspond to two different energy sources: the circularization luminosity, and the accretion luminosity.  The former evolves on the fallback timescale $t_{\rm f}$ while the latter evolves on a viscous timescale, 
\begin{equation}
t_{\rm v} = \sqrt{\frac{8R_{\rm p}^3}{GM_\bullet}}\alpha^{-1} \left[\frac{H(2R_{\rm p})}{2R_{\rm p}} \right]^{-2}
\end{equation}
where we have assumed that the disk outer edge is $2R_{\rm p}$, $\alpha<1$ is the usual Shakura-Sunyaev viscosity parameter, and $H/R$ is the disk aspect ratio.  For standard TDEs, these two sources of luminosity are difficult to disentangle, because $t_{\rm v} \ll t_{\rm f}$ \cite{Ulmer99, StrubbeQuataert09}: as soon as matter circularizes into a disk, it drains rapidly into the SMBH, and the two sources of luminosity rise and fall  together.  

However, for very massive SMBHs, $\dot{M}_{\rm p}<\dot{M}_{\rm Edd}$ and the accretion disk will be geometrically thin, dramatically increasing $t_{\rm v}$ and producing an inverted timescale hierarchy: $t_{\rm v} \gtrsim t_{\rm f}$.  The circularization luminosity can be released promptly, but accretion luminosity will be bottlenecked by the long $t_{\rm v}$.  
The range of inferred SMBH masses for ASASSN-15lh yield $\dot{M}_{\rm p} \lesssim \dot{M}_{\rm Edd}$ (unlike for most other TDE hosts),  so it is natural to expect two power sources - circularization and accretion - to have two different peaks.

The characteristic decay time of the first peak will be $\sim t_{\rm f}$, while that of the second will be $t_{\rm v}$, which is a sensitive function of $\alpha$ and $H/R$.  Plausible parameter choices (e.g. $M_\bullet = 10^8 M_\odot$, $\alpha = 0.1$, $H/R = 0.03$) give $t_{\rm v}\sim 6~{\rm months}$, in agreement with the observed second component of the light curve.  However, the first component of the light curve decays on a timescale of $\sim 1-2~{\rm month}$, unlike the Newtonian estimate of equation \ref{eq:tFall}: $t_{\rm f} \approx 1 ~{\rm yr}$.  In the following subsection, we argue that general relativity effects can increase $\Delta \epsilon$ in TDEs with extremely relativistic pericenters, substantially reducing $t_{\rm f}$ for this subset of TDEs.

\subsection{Relativistic Alterations to the Fallback Time}

Highly relativistic tidal disruption will alter the Newtonian  $\Delta \epsilon$.  The frozen-in energy spread $\Delta \epsilon$ can increase by a factor up to $\sqrt{2}$ (ref.~\cite{Kesden12b}), decreasing $t_{\rm f}$ and increasing $\dot{M}_{\rm p}$ modestly.  However, $\Delta \epsilon$ may increase further by modest spin-orbit misalignments in a TDE for sufficiently relativistic pericenters \cite{Stone13}.  Here, we argue that this effect can strongly reduce the fallback time from the Newtonian estimate of equation \ref{eq:tFall}.

The extremity of a TDE is quantified not just by $B\equiv R_{\rm p}/R_{\rm g}$, but also by the penetration factor $\beta \equiv R_{\rm t}/R_{\rm p}$.  TDEs with a large $\beta$  will strongly compress the star orthogonal to the orbital plane, causing a vertical collapse with velocity $v_{\rm z} \approx \beta \sqrt{GM_\star / R_\star}$.  The collapse is reversed near pericenter, once internal pressure in the disrupted star builds up to the point where it ``bounces'' along this vertical axis\cite{CarterLuminet83}, receiving an almost impulsive hydrodynamic kick $\approx v_{\rm z}$ along the direction of collapse.  The bounce typically has little effect on the energy spread of the debris\cite{Guillochon13, Hayasaki+13}, despite the fact that $V_{\rm p} v_{\rm z} \gtrsim \Delta\epsilon$ for large $\beta$ or small $B$ (here $V_{\rm p}$ is the orbital velocity at pericenter).  One could define a hydrodynamic component of the energy spread, $\Delta\epsilon_{\rm h} \equiv \vec{V}_{\rm p} \cdot \vec{v}_{\rm z} = V_{\rm p}v_{\rm z} \cos \theta$, but in standard TDEs the misalignment angle $\theta=\pi/2$.

However, for very low $B$, modest spin-orbit misalignment will precess the orbital plane as the star passes through the tidal sphere, partially aligning the axis of vertical collapse with the orbital velocity vector.  The per-orbit precession in the line of ascending nodes for a parabolic orbit is\cite{Merritt+10}:
\begin{equation}
\delta \Omega = \sqrt{2}\pi a_\bullet \left(\frac{R_{\rm g}}{R_{\rm p}} \right)^{3/2}
\end{equation}
at leading post-Newtonian order.  Since most of this shift occurs near pericenter, we approximate the rotation in the line of ascending nodes between $R_{\rm t}$ and the bounce (near $R_{\rm p}$) as $\delta\Omega / 2$.  The misalignment angle is $\theta$ ($\cos \theta \approx \frac{\delta\Omega}{2}\sin I$), where $I$ is the misalignment between spin and orbital angular momentum.  Defining a total energy spread $\Delta \epsilon_{\rm tot} \equiv \Delta \epsilon + \Delta \epsilon_{\rm h}$, we find (Stone, Kennon, \& Metzger, manuscript in preparation):
\begin{equation}
\Delta \epsilon_{\rm tot} = \Delta \epsilon(1+a_\bullet \beta^{3/2}B^{-3/2} \sin I)
\end{equation}

The above arguments are approximate, as the post-Newtonian approximation begins to break down for $R_{\rm p} \sim R_{\rm g}$, and we have treated the bounce hydrodynamics impulsively.  A detailed examination of this spin-orbit coupling is beyond the scope of this paper, but order unity increases in $\Delta \epsilon$ and decreases in $t_{\rm f} \propto \Delta\epsilon^{3/2}$ are expected for TDEs around SMBHs with $a_\bullet \approx 1$ and $B \sim 1$.

\subsection{Relativistic Calculation of the Hills Mass}
Some of the previous Newtonian estimates break down when the star's orbital pericenter $R_{\rm p} \sim R_{\rm g}$.  
In this regime, general relativistic effects are crucial for proper modelling of the Hills mass\cite{Beloborodov92, Kesden12a}.  The increased $M_{\rm H}$ arises from two different effects: first, a large $a_\bullet$ will lower the IBCO radius significantly, and second, the Kerr tidal field is somewhat stronger than the Schwarzschild equivalent.

To calculate $M_{\rm H}(a_\bullet)$, we employ Fermi normal coordinates to write a local, fully general relativistic tidal tensor\cite{Marck83}.  We then employ the accepted formalism\cite{Kesden12a} to estimate relativistic Hills masses.  The results are shown for equatorial orbits (that maximize $M_{\rm H}$) in Figure \ref{fig:HillsMass}, where we see that the relativistic Kerr $M_{\rm H}$ can be almost an order of magnitude greater than the Newtonian (or Schwarzschild) equivalent.

\subsection{Data Availability Statement}
The photometry and spectra of ASASSN-15lh supporting the findings of this study are available from WISeREP\cite{WISeREP} ($\rm{http://wiserep.weizmann.ac.il/}$)

\end{methods}



\clearpage

\begin{addendum}

 \item[Supplementary Information]

\begin{SIfigure}
\centering
\includegraphics[width=16cm]{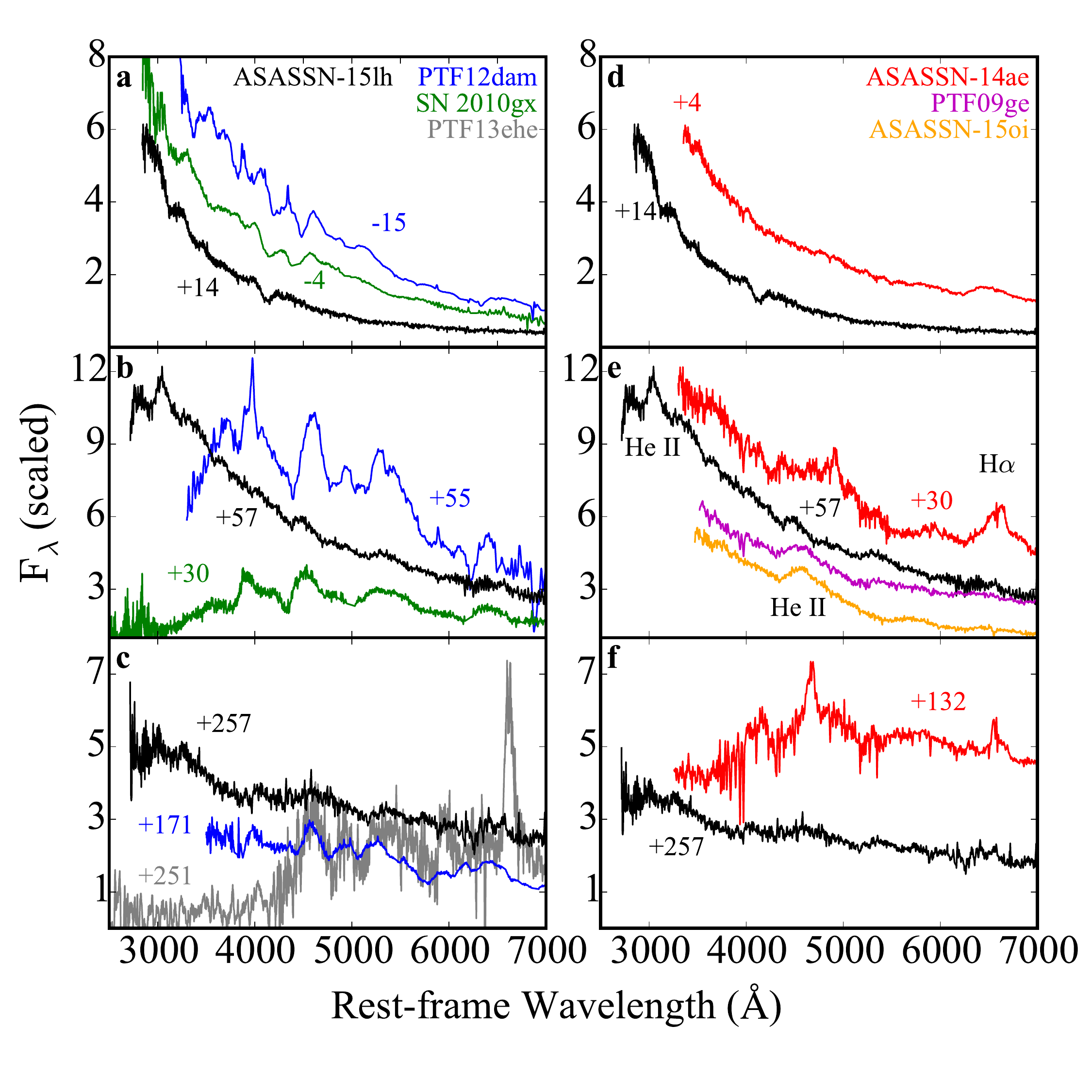}
\caption{\boldmath$\vert$\unboldmath \hspace{1em}  \textbf{
Spectroscopic comparison of ASASSN-15lh with SLSNe and TDEs}.
The left panels compare spectra of ASASSN-15lh at different phases with those of SLSNe\cite{Pastorello10gx,Nicholl12dam,Yan13ehe}.
The right panels compare the same spectra with TDEs\cite{Arcavi14,Holoien14ae,Holoien15oi}. 
At early times (a), ASASSN-15lh looks indeed similar to SLSNe\cite{Dong16} but lacks a strong O \textsc{ii} feature.
The later spectra (b,c) are very different from those of SLSNe and even at +257 days ASASSN-15lh is purely continuum-dominated without any nebular features or signs of strong circumstellar interaction, such as in PTF13ehe\cite{Yan13ehe}.   
Despite the differences with TDE spectra (d,e,f) there are also similarities, especially with TDEs showing blueshifted He \textsc{ii} (e), or even with the early spectrum of ASASSN-14ae (d). ASASSN-15lh has H (Figure~\ref{fig:specseq}), but it is weaker than in ASASSN-14ae.
\label{fig:specComp}
}
\end{SIfigure}

\newpage

\begin{SIfigure}
\centering
\includegraphics[width=16cm]{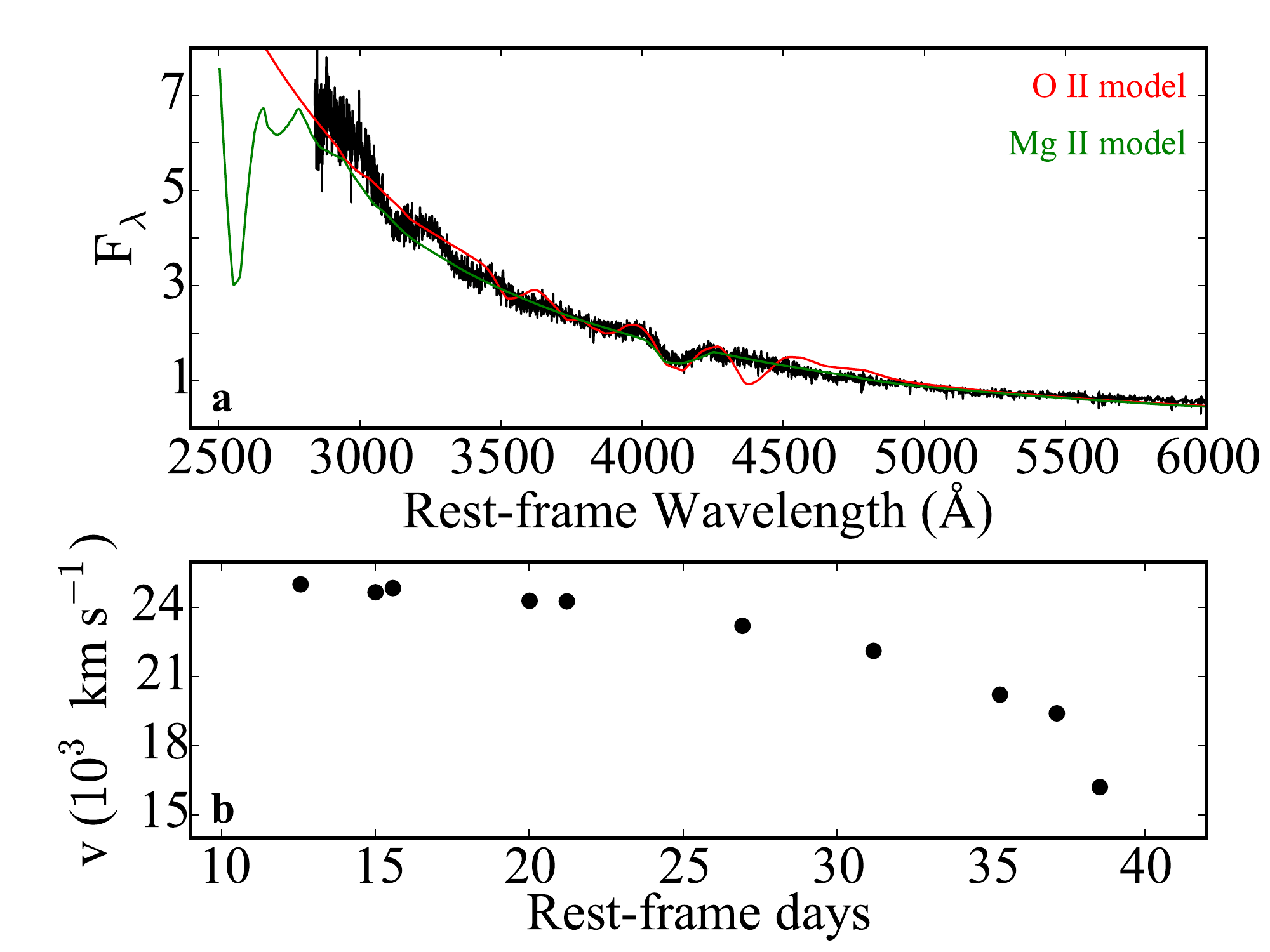}
\caption{\boldmath$\vert$\unboldmath \hspace{1em}  \textbf{The early phases of ASASSN-15lh.} (a) Modelling the spectra with SYNOW\cite{Hatano99} and O \textsc{ii} as in SLSNe\cite{Quimby11,Mazzali16} is not possible as a strong feature at $\sim$4400 \AA\ is inevitable. 
A tentative identification of the strongest feature at $\sim$4100 \AA\ is Mg \textsc{ii}, which however produces additional strong features in the UV.  
Such features (transient in nature) have been observed in a TDE candidate\cite{2014ApJ...780...44C} and cannot be ruled out for the early phases of ASASSN-15lh, based on the available spectra.
The use of SYNOW in this case is purely illustrative, as this code has not been made to model TDEs. 
(b) Velocity evolution of the absorption at $\sim$4100 \AA\ assuming it is Mg~\textsc{ii}.
Another possibility for this feature is that it might be due to high-velocity He \textsc{ii}\cite{StrubbeQuataert11}. In this case, the velocities are larger by $\sim$12,000 km s$^{-1}$.
\label{fig:SynowVels}
}
\end{SIfigure}

\newpage

\begin{SIfigure}
\centering
\includegraphics[width=16cm]{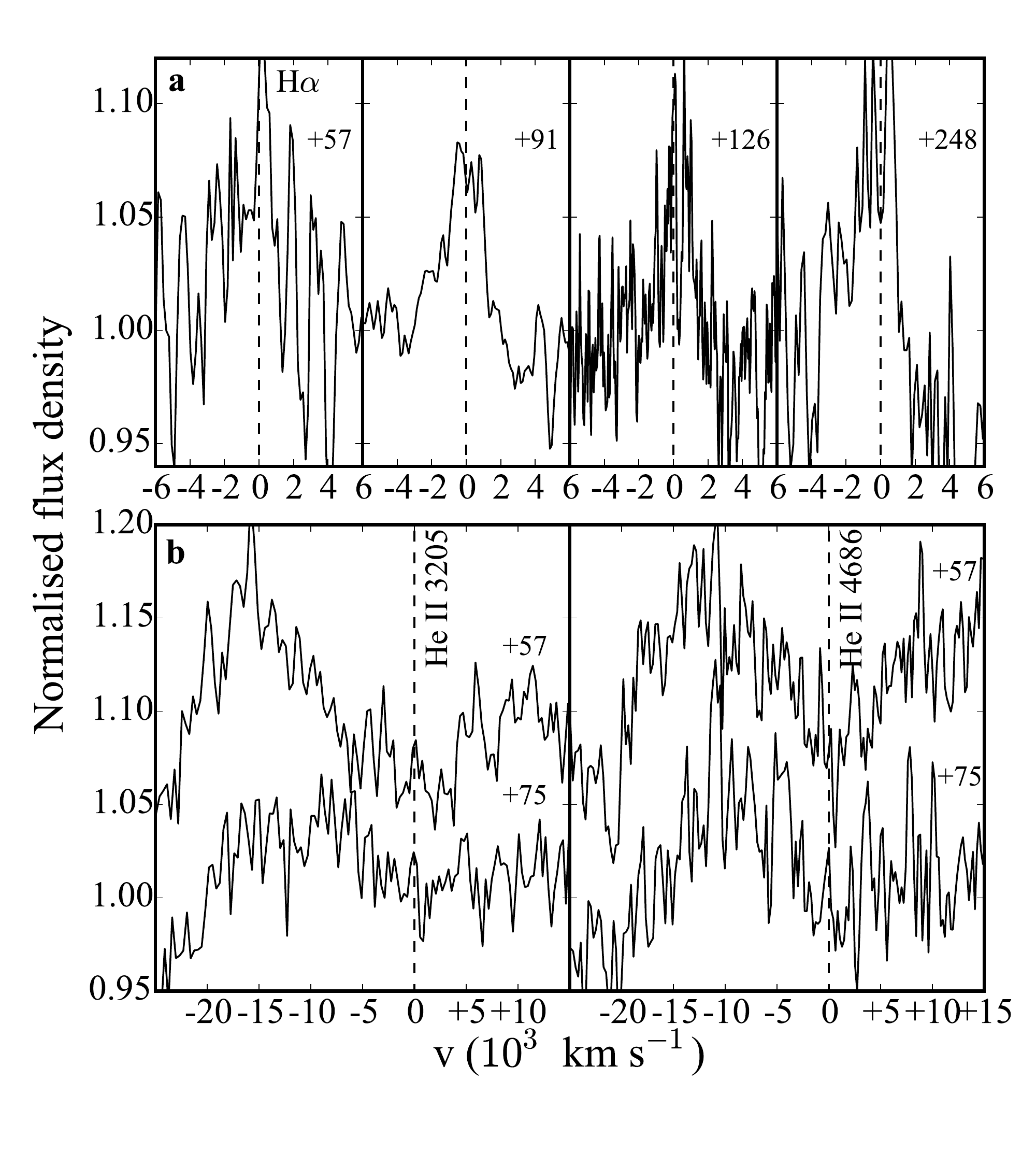}
\caption{\boldmath$\vert$\unboldmath \hspace{1em}  \textbf{Detection of H and possible detection of transient He in the spectra of ASASSN-15lh. }
(a) H$\alpha$ at representative epochs. The line is weak and the detection significance varies with signal-to-noise (see also Figure \ref{fig:specseq}). 
However, there is no measurable evolution in its strength (EW $\sim 4-8$ \AA), and its presence cannot be excluded in any spectrum.
(b) The profiles of the lines identified as blueshifted He~\textsc{ii} at $+57$ and $+75$ days.
\label{fig:HeII}
}
\end{SIfigure}

\clearpage

\begin{SIfigure}
\centering
\includegraphics[width=16cm]{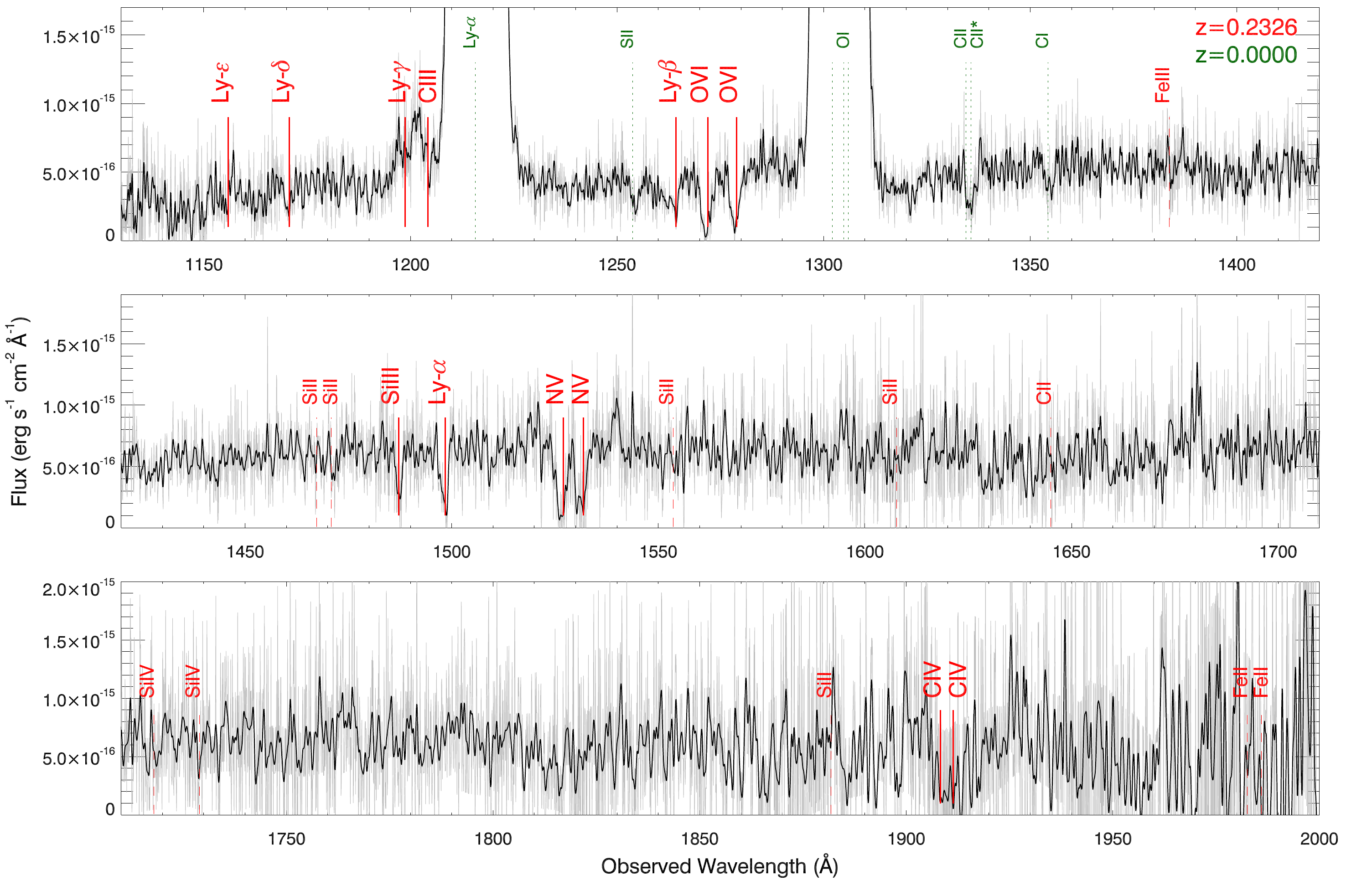}
\caption{\boldmath$\vert$\unboldmath \hspace{1em}  \textbf{An ultraviolet spectrum of ASASSN-15lh obtained with \textit{HST}+COS at 168 rest-frame days past maximum.}
We identify both local geocoronal lines (marked with green) and narrow absorption features associated with ASASSN-15lh at $z=0.2326$ (red).
Detections (marked with red solid lines), include  Ly-$\alpha$, Ly-$\beta$, Si \textsc{iii}, C \textsc{iv}, and, notably, N \textsc{v} and O \textsc{vi}. 
The most prominent lines that are not detected are marked with a red dashed line.
\label{fig:COS}
}
\end{SIfigure}

\newpage

\begin{SIfigure}
\centering
\includegraphics[width=16cm]{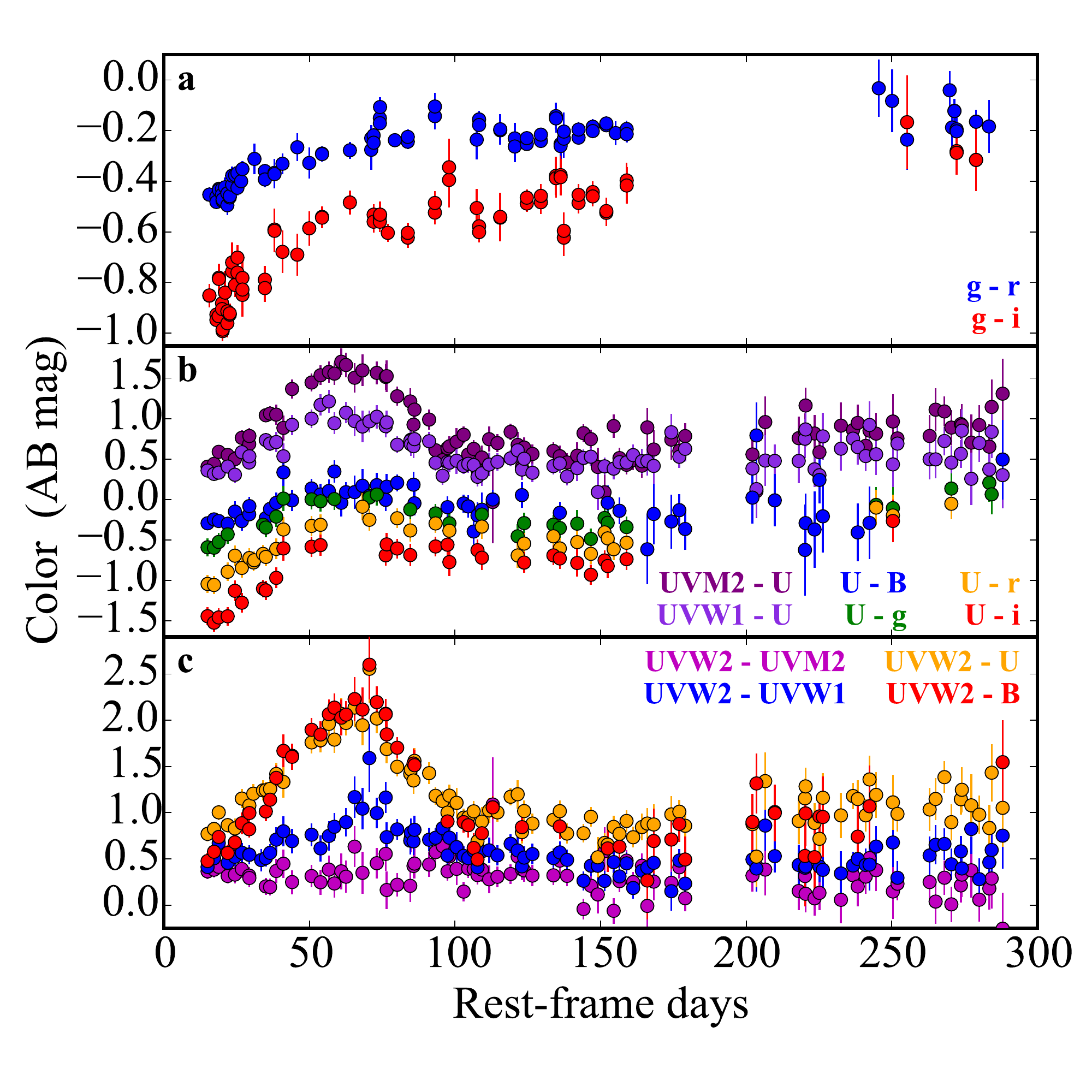}
\caption{\boldmath$\vert$\unboldmath \hspace{1em}  \textbf{The colour evolution of ASASSN-15lh. }
We show optical colours with respect to $g$ in panel (a), different UV and UV$-$optical colours with respect to $U$ in panel (b), and UV colours with respect to $UVW2$ in panel (c).
Errorbars represent 1$\sigma$ uncertainties. Most colours show a significant evolution to the red, peaking at 60 days past-maximum and corresponding to the UV minimum in Figure~\ref{fig:LC}. Subsequently, the UV colours get bluer again. All colours remain fairly constant after day $+100$ and for a period of over 120 days.
\label{fig:colors}
}
\end{SIfigure}

\clearpage

\begin{SIfigure}
\centering
\includegraphics[width=16cm]{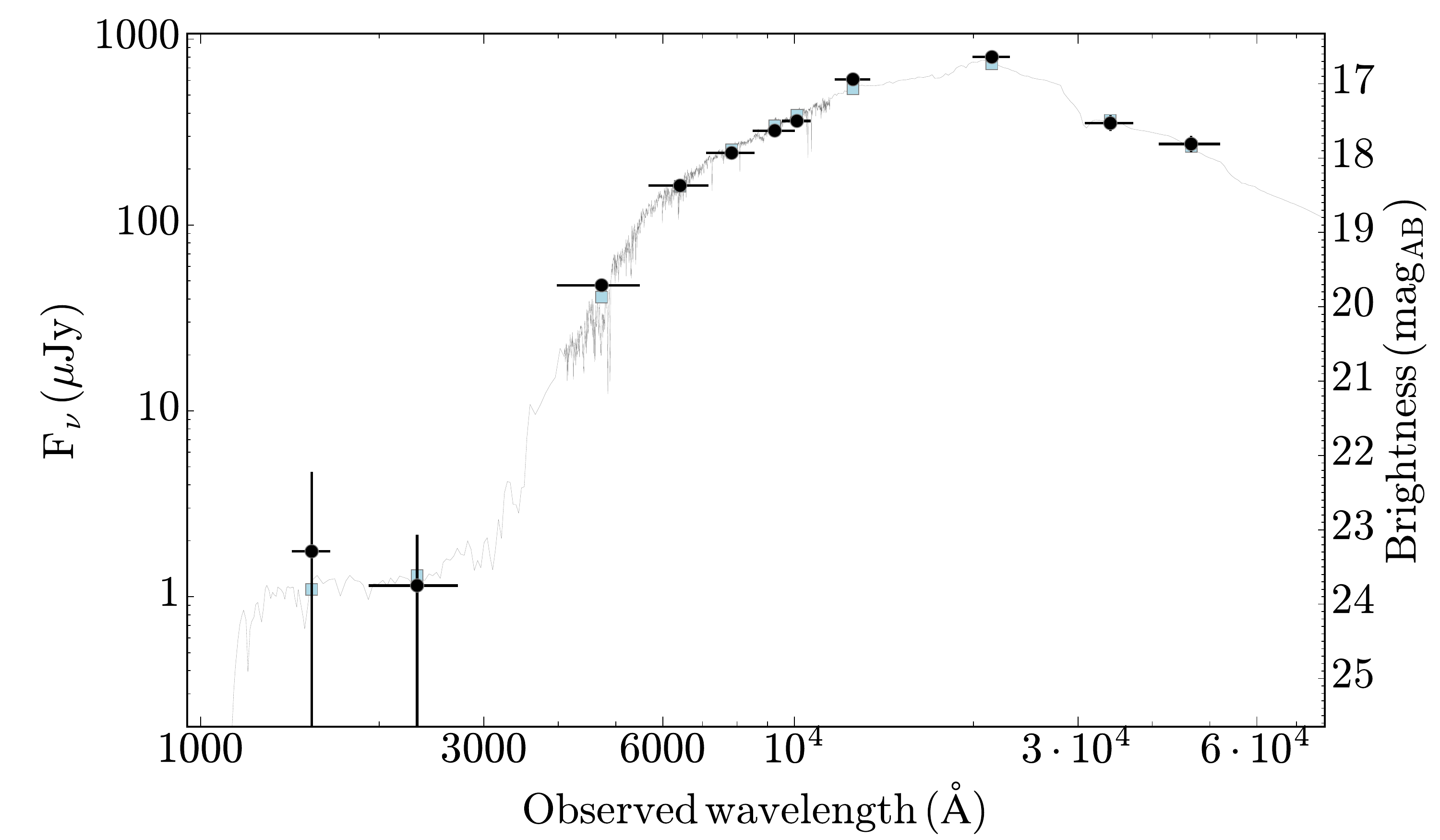}
\caption{\boldmath$\vert$\unboldmath \hspace{1em}  \textbf{SED fit for the host galaxy of ASASSN-15lh. }
To perform the fit we are using the code \textit{Le Phare} and photometric data all the way from the far-UV (GALEX) to the mid-IR (WISE). 
Shown is the best-fitting template (reduced $\chi^2 = 0.998$). 
The GALEX non-detections are shown here with the nominal SExtractor photometry errors.
\label{fig:SEDfit}
}
\end{SIfigure}

\newpage

\begin{SIfigure}
\centering
\includegraphics[width=16cm]{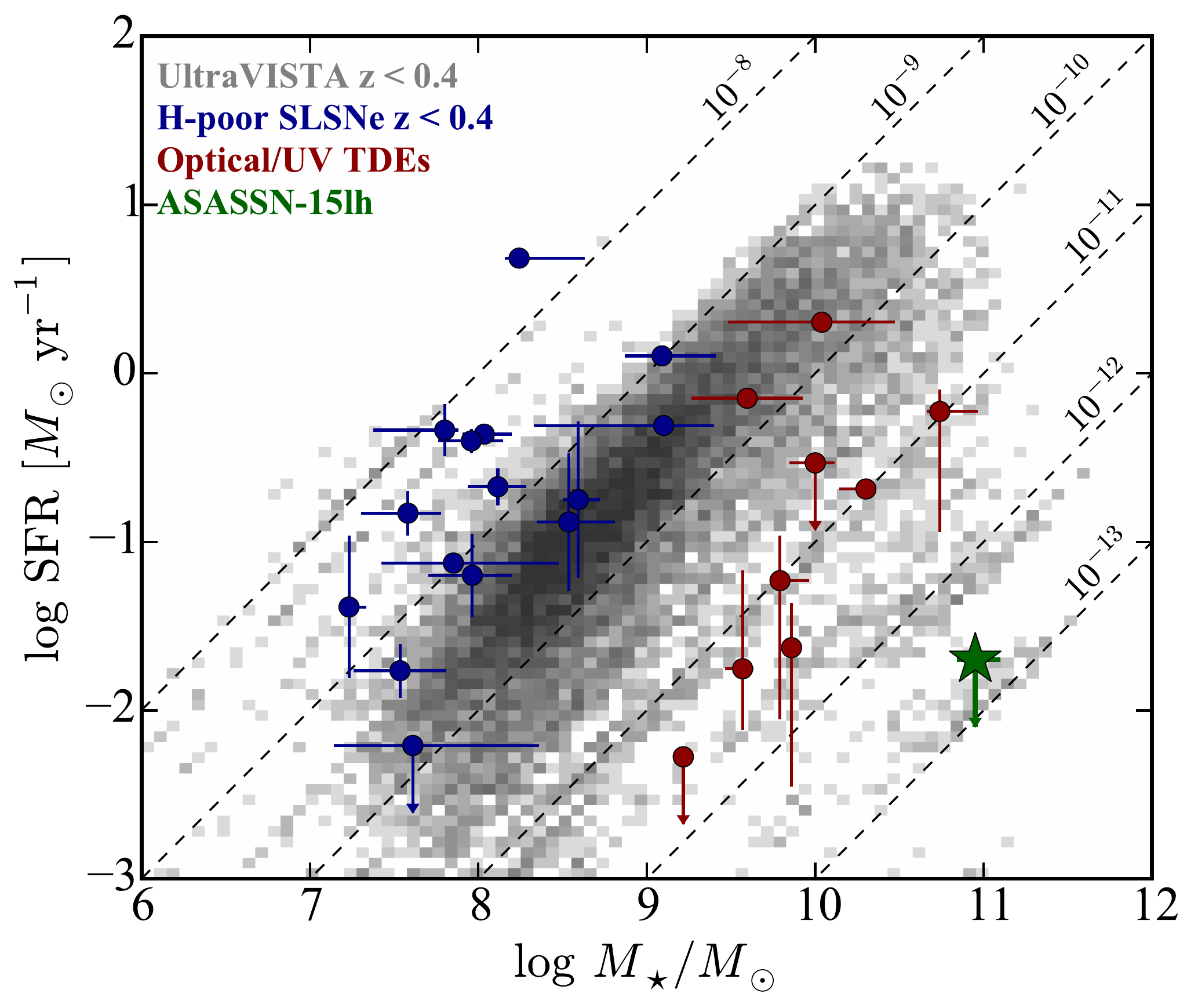}
\caption{\boldmath$\vert$\unboldmath \hspace{1em}  \textbf{The properties of the host galaxy of ASASSN-15lh compared to those of SLSNe\cite{Lunnan14,Leloudas15} and TDEs\cite{Arcavi14,Holoien14ae,Holoien14li,Holoien15oi}.}
For comparison, we show the general galaxy population at $z<0.4$ from UltraVISTA and we have drawn lines of equal specific SFR.
The location of ASASSN-15lh clearly stands out from those of SLSNe that are found in dwarf star-forming galaxies on or above the main sequence of star formation. The separation of ASASSN-15lh is 3 orders of magnitude in terms of specific SFR.
At the same time, the host is significantly more massive from those of most optical TDEs, suggesting a large SMBH mass. 
\label{fig:hostsComp}
}
\end{SIfigure}

\newpage

\begin{SIfigure}
\centering
\includegraphics[width=16cm]{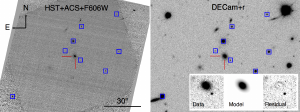}
\caption{
\boldmath$\vert$\unboldmath \hspace{1em}  \textbf{The localisation of ASASSN-15lh in the nucleus of its host galaxy.}
On the left is an \textit{HST} image of ASASSN-15lh obtained at day +47.
On the right is a DECam image from October 2014, several months before the transient appeared.
The sources used to derive the geometric transformation between the frames are marked with blue squares. 
The inset shows a zoom in to the host together with a {\sc galfit} model and the residual from the model fit.
ASASSN-15lh has an offset of $36 \pm 53$ mas  from the nucleus and it is thus consistent with the location of the central SMBH.
\label{fig:astrom}}
\end{SIfigure}


\clearpage

\begin{SItable}
\begin{center}
\begin{tabular}{|l|c|c|c|c|c|}
\hline
Line				& $\lambda_{obs}$ (\AA)  & offset (km/s) $^a$   &	EW	(\AA)		&EW$_{GRB}$ $^b$  (\AA)  &  EW$_{SLSN}$ (\AA)\\ 
\hline
O \textsc{vi} 1031	&	1271.57		& $-$80 $\pm$ 12	&		1.97	$\pm$ 0.26	&  		--     	 	&		--	     	\\  
O \textsc{vi} 1037	&	1278.55		& $-$80 $\pm$ 12	&		1.57	$\pm$ 0.24	&  		--     	 	&		--	     	\\  
S \textsc{iii} 1206	&	1487.26		& $+$44 $\pm$ 22	&		0.79	$\pm$ 0.26	&  		--     	 	&		--	     	\\  
Ly-$\alpha$		&	1498.26		&	0$  \pm$ 12	&		1.15	$\pm$ 0.32	&  		73.0     	 &		--    	  	\\ 
N \textsc{v} 1238	&	1526.50		&$-$117 $\pm$ 12	&		1.97	$\pm$ 0.41	&            	0.14 		&		--	    	 \\ 
N \textsc{v} 1242	&	1531.26		&$-$117 $\pm$ 12	&		2.07	$\pm$ 0.43	& 		0.07  	 &		--	     	\\ 
Si \textsc{ii} 1259	&	--			&		--		&		$< 1.73$			&  	        1.26	     	 &		--	     	\\
Si \textsc{ii} 1304	&	--			&	--			&		$< 2.21$			&  	        2.29	     	 &		--	     	\\
C \textsc{ii} 1334	&	--			&	--			&		$< 2.56$			&  	        1.73	     	 &		--	     	\\  
Si \textsc{iv} 1393	&	--			&	--			&		$< 2.75$			&  	        0.95	     	 &		--	     	\\  
Si \textsc{ii} 1527	&	--			&	--			&		$< 6.7$			&  	        0.93	     	 &		--	     	\\
C \textsc{iv}/C \textsc{iv} 1548	&	--	&$-$204 $\pm$ 92	&	      3.08 $\pm$ 2.35 		 &              2.18		 &              --                \\
Fe \textsc{ii} 1608	&	--			&	--			&		$<$ 24.0			&  		0.85     	 &		--	     	\\
Fe \textsc{ii} 2382	&	--			&	--			&		$<$ 1.31			&  		1.65     	 & 0.35 $\pm$0.03 $^c$   \\  
Fe \textsc{ii} 2600	&	--			&	-- 			&		$<$ 3.65			&  		1.85     	 & 0.29 $\pm$0.03 $^c$   \\  
Mg \textsc{ii} 2796	&	3445.67		&  $+$8  $\pm$ 8	&		0.50	$\pm$ 0.05	&  		1.71  	 &	2.6$\pm$ 1.2 $^d$ \\  
Mg \textsc{ii} 2803	&	3454.46		& $+$14 $\pm$10	&		0.38	$\pm$ 0.05	&  		1.47  	 &	 --   $^d$    		 \\  
\hline
\end{tabular}
\end{center}
\caption{
\boldmath$\vert$\unboldmath \hspace{1em} \textbf{Absorption lines in the spectrum of ASASSN-15lh}. Upper limits are 3$\sigma$. EWs are in rest-frame. Notes: $^a$ The reference velocity was set to Ly-alpha ($z=0.23253$). $^b$ Based on a high S/N composite GRB afterglow spectrum\cite{Christensen2011}. Typical error for weak lines is 0.02~\AA. $^c$ Based only on PTF13ajg\cite{Vreeswijk14}. $^d$ Total value for the doublet, based on a sample of 13 events\cite{Vreeswijk14}.
\label{tab:abslines}}
\end{SItable}

\clearpage

\begin{SItable}
\begin{center}
\begin{tabular}{|c|c|c|c|c|}
\hline
Date (UT)         & Phase	& Telescope+Instrument     & Grism            & Range (nm)    \\ 
\hline
2015-06-22.7    &	$+$14.0	& FTS+FLOYDS               &            			&   325 -  930        \\
2015-06-24.7    & 	$+$15.6	& ANU 2.3m+WiFeS    	& B3000+R3000    	& 350 - 956    \\
2015-07-01.6    & 	$+$21.2	& ANU 2.3m+WiFeS    	& B3000+R3000    	& 350 - 956    \\
2015-07-08.7    & 	$+$26.9	& ANU 2.3m+WiFeS    	& B3000+R3000    	& 350 - 956    \\
2015-07-21.3    & 	$+$37.1	& VLT+FORS2        		& 300V            		& 445 - 865    \\
2015-07-31.5    & 	$+$45.4	& ANU 2.3m+WiFeS    	& B3000+R3000    	& 350 - 956    \\
2015-08-14.3    & 	$+$56.6	& NTT+EFOSC2    		& GR\#11+16        	& 334 - 999    \\
2015-09-06.0    & 	$+$75.0	& NTT+EFOSC2    		& GR\#11+16        	& 334 - 999    \\
2015-09-16.1    & 	$+$83.2	&  NTT+EFOSC2    $^a$	& GR\#11+16        	& 334 - 999    \\
2015-09-23.2    & 	$+$89.0	& NTT+EFOSC2    		& GR\#11+16        	& 334 - 999    \\
2015-09-25.1    & 	$+$90.6	& VLT+FORS2        		& 300V            		& 445 - 865    \\
2015-10-11.1    & 	$+$103.6	&  NTT+EFOSC2   $^b$	& GR\#11            	& 334 - 745    \\
2015-10-13.2    & 	$+$105.2	& NTT+EFOSC2    $^b$	& GR\#16            	& 599 - 999    \\
2015-11-08.0    & 	$+$126.2	& Magellan+IMACS  		&    Gri-300-17.5       &   400-999         \\
2015-11-18.1    & 	$+$134.3	& NTT+EFOSC2    		& GR\#13            	& 365 - 924    \\
2015-12-16.0    & 	$+$157.0	& NTT+EFOSC2    		& GR\#11+16        	& 334 - 999    \\
2016-04-06.3    & 	$+$248.1	& NTT+EFOSC2    		& GR\#11+16        	& 334 - 999    \\
2016-04-15.3    & 	$+$256.2	& NTT+EFOSC2    		& GR\#11+16        	& 334 - 999    \\
\hline
\end{tabular}
\end{center}
\caption{\boldmath$\vert$\unboldmath \hspace{1em} \textbf{Log of spectra}. Notes: $^a$ Low S/N; not used. $^b$ Combined to a single spectrum.
\label{tab:speclog}}
\end{SItable}

\end{addendum}

\end{document}